\documentclass[12pt]{iopart}

\usepackage{graphicx}
\usepackage{dcolumn}
\usepackage{bm, amssymb, color}
\usepackage{hyperref}
\usepackage{epstopdf}
\usepackage{float}
\usepackage{soul}

\usepackage[textwidth=16cm, top=1in, bottom=1in]{geometry}
\usepackage{extsizes}

\usepackage{amssymb} 
\usepackage{bm} 

\begin{document}

\title{Exclusion Volumes of Convex Bodies in High  Space Dimensions: Applications to Virial Coefficients and Continuum Percolation}

\author{Salvatore Torquato$^{1,2,3,4,*}$ and Yang Jiao$^{5,6}$ }

\address{$^1$
Department of Chemistry, Princeton University, Princeton, New Jersey 08544, USA}
\address{$^2$
Department of Physics,  Princeton University, Princeton, New Jersey 08544, USA}
\address{$^3$
Princeton Institute  of Materials, Princeton University, Princeton, New Jersey 08544, USA}
\address{$^4$
Program in Applied and Computational Mathematics, Princeton
University, Princeton, New Jersey 08544, USA}
\ead{torquato@electron.princeton.edu}

\address{$^5$
Materials Science and Engineering, Arizona State University,
Tempe, Arizona 85287, USA}
\address{$^6$
Department of Physics, Arizona State University, Tempe, Arizona
85287, USA}
%

\date{\today}
\begin{abstract}
Using the concepts of mixed volumes and
quermassintegrals of convex geometry, we derive an exact formula for the exclusion volume $v_{\mbox{\scriptsize ex}}(K)$ for a general convex body $K$
that applies in any space dimension.
While our main interests concern the rotationally-averaged exclusion volume of a convex body with respect to another convex body,
we also describe some results for the exclusion volumes for convex bodies with the same orientation.
We show that the sphere minimizes the dimensionless exclusion volume $v_{\mbox{\scriptsize ex}}(K)/v(K)$
among all convex bodies, whether randomly oriented or uniformly oriented, for any $d$, where $v(K)$ is the volume of $K$.
When the bodies have the same orientation, the simplex maximizes the dimensionless exclusion volume for any $d$ with a large-$d$ asymptotic scaling behavior of $2^{2d}/d^{3/2}$,
which is to be contrasted with the corresponding scaling of $2^d$ for the sphere. We present explicit formulas for  quermassintegrals $W_0(K),\ldots, W_d(K)$ for  many different nonspherical convex bodies,
including cubes, parallelepipeds, regular simplices, cross-polytopes,
cylinders, spherocylinders, ellipsoids as well as lower-dimensional bodies, such as hyperplates and line segments. These results are utilized
 to determine  the rotationally-averaged exclusion volume $v_{\mbox{\scriptsize ex}}(K)$  for these convex-body shapes for dimensions 2 through 12. While the sphere is the shape possessing the minimal dimensionless exclusion volume, we show that, among the convex bodies considered that are sufficiently compact, the simplex possesses the maximal $v_{\mbox{\scriptsize ex}}(K)/v(K)$ with a scaling behavior of $2^{1.6618\ldots d}$. Subsequently,  we apply these results to determine the corresponding second virial coefficient $B_2(K)$ of the aforementioned  hard hyperparticles. Our results are also applied to compute estimates of the continuum percolation threshold
$\eta_c$ derived previously by the authors for systems of identical  overlapping convex bodies. We
conjecture that overlapping spheres possess the maximal value of $\eta_c$ among all
identical nonzero-volume convex overlapping bodies for $d \ge 2$, randomly or uniformly oriented,  and that, among all identical, oriented nonzero-volume
convex bodies, overlapping simplices have the minimal value of $\eta_c$ for $d\ge 2$.

\end{abstract}
\maketitle



\section{Introduction}

\begin{figure}[ht]
\begin{center}
\includegraphics[height=2in,keepaspectratio,clip=]{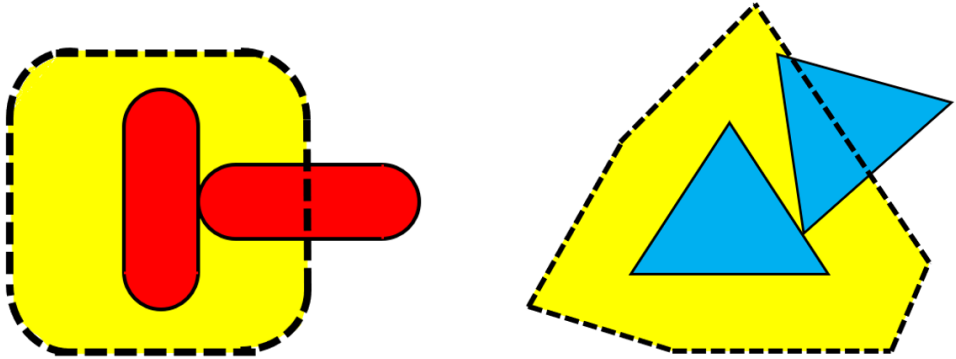}
\end{center}
\caption{(Color online) Illustration of the exclusion volume of two-dimensional  convex bodies with a fixed orientation
with respect to one another. In each of the two examples, the exclusion volume is the region interior to the boundary delineated  by the dashed lines.
Left panel: Centrally symmetric spherocylinder.  Right panel: Non-centrally symmetric equilateral triangle (two-dimensional
regular simplex).}
\label{exclusion-1}
\end{figure}

The exclusion volume of an arbitrary $d$-dimensional convex body (hyperparticle) $K$ in $d$-dimensional Euclidean space $\mathbb{R}^d$ is a fundamental quantity that arises
in the virial expansion of the pressure of a hard-hyperparticle fluid \cite{Kih53,Lu82,Ha86,Tar91}, estimates of
continuum percolation thresholds \cite{Ba84,Bu85,To12a,To12c,To13a} and a variety of problems involving the distance of closest approach
of two nonoverlapping bodies  \cite{On44,On49,Fr87,Pe85,Do05b,To09b}. The exclusion volume is the region of space that is excluded to one hyperparticle
due the presence of another one with a specific relative orientation. Whereas in the case of hyperspheres or {\it oriented} centrally-symmetric hyperparticles
(e.g, cubes or cross-polytopes), the exclusion volume relative to the volume of the convex body $K$, $v(K)$, is simply equal to $2^d$, its determination
for general convex bodies with arbitrary relative orientations is nontrivial.
The generalized exclusion volume $v_{\mbox{\scriptsize ex}}(K)$ of a convex body $K$
can be expressed as
\begin{equation}
\label{v_ave}
v_{\mbox{\scriptsize ex}}(K)=\int_{\mathbb{R}^d} f({\bf r},{\boldsymbol \omega};K) p({\boldsymbol \omega};K) d{\bf r}d{\boldsymbol \omega},
\end{equation}
where $ f({\bf r},{\boldsymbol \omega};K)$ is the exclusion-region
indicator function \cite{To12a,To13a}, ${\bf r}$ and ${\boldsymbol \omega}$ is the
centroid position and orientation of one body, respectively,
with respect to a coordinate system at the centroid of the other
body with some fixed orientation, and $p({\boldsymbol \omega};K)$ is the orientational probability density function.
Figure \ref{exclusion-1} provides two-dimensional examples of exclusion volumes for a centrally symmetric
body and noncentrally symmetric body.

In this paper, we derive exact general expressions for the
exclusion volume  $v_{\mbox{\scriptsize ex}}(K)$ of a  convex body $K$  for any dimension in terms of
of the quermassintegrals and mixed volumes of convex geometry \cite{Sa76,St95,Sc14,Me20}. While we mainly treat
the rotationally-averaged exclusion volume of a convex body with respect to another convex body, we also describe some results for the exclusion volumes for convex bodies with the same orientation.
We then provide explicit formulas for quermassintegrals for specific nonspherical convex bodies, including cubes,  cylinders, spherocylinders, parallelepipeds, regular simplices, cross-polytopes,
 ellipsoids as well as lower-dimensional bodies, such as hyperplates and line segments.  These results are employed
 to determine  the rotationally-averaged exclusion volume for these specific convex-body shapes for dimensions 2 through 12. Note that
the three regular polytopes considered here (cube, regular simplex
and cross-polytope) are the only regular polytopes possible for $d \ge 5$ \cite{Co73}.

Subsequently, we apply these results to determine  the corresponding second virial coefficient $B_2(K)$ that arises
in the virial expansion of the pressure $P$ of hard-particle fluids composed of such shapes through second order in the density \cite{Ha86}, namely,
\begin{equation}
\frac{P}{k_B T}= \rho + \rho^2 B_2(K) +{\cal O}(\rho^3),
\label{virial}
\end{equation}
where $\rho$ is the number density, i.e., the number of convex bodies per unit volume in the thermodynamic limit,
$k_B$ is the Boltzmann constant, $T$ is absolute temperature, and
\begin{equation}
B_2(K)=\frac{1}{2} v_{\mbox{\scriptsize ex}}(K).
\label{eq_b2_vex}
\end{equation}
It is noteworthy that third- and higher-order terms in the viral expansion (\ref{virial}) of
a hard-particle fluid become negligibly small
in the high-$d$ asymptotic limit \cite{Fr99}. The exclusion volume arises at the second virial level due to the nonoverlap constraint of two convex bodies
when they come in contact with one another. While $B_2(K)$ has long been known for a variety of nonspherical hard
particles in two and three dimensions \cite{Kih53}, no such results appear to have been obtained for $d \ge 4$ until the present work.
We shall also apply our results to compute estimates of the continuum percolation threshold
$\eta_c$ derived by Torquato and Jiao \cite{To13a} for systems of identical {\it overlapping} convex bodies with arbitrary
orientational distributions in $d$-dimensional Euclidean space $\mathbb{R}^d$ that depend
on the  exclusion volume $v_{\mbox{\scriptsize ex}}(K)$, which arises here
because it defines the region when two hyperparticles overlap  \cite{To13a}. In particular, they derived
the following lower bound on the percolation threshold $\eta_c$ that is applicable for any $d$:
\begin{equation}
\eta_c \ge \frac{v(K)}{v_{\mbox{\scriptsize ex}}(K)},
\label{bound}
\end{equation}
where $\eta=\rho v(K)$ is a dimensionless (reduced) density.
The inequality (\ref{bound}) applies in all dimensions, becomes sharper as dimension increases, and exact in the limit
$d \to \infty$ \cite{To13a}.
Torquato and Jiao \cite{To13a} conjectured the following sharper scaling relation to estimate $\eta_c$:
\begin{equation}
    \eta_c \approx \frac{(\eta_c)_S}{(v/ v_{\mbox{\scriptsize ex}})_S}\left(\frac{v(K)}{v_{\mbox{\scriptsize ex}}(K)}\right) = 2^d\left(\frac{v(K)}{v_{\mbox{\scriptsize ex}}(K)}\right)(\eta_c)_S,
\label{scaling-relation}
\end{equation}
where $(\eta_c)_S$  and $(v_{\mbox{\scriptsize ex}}/v)_S = 2^d$
are the percolation threshold and dimensionless exclusion volume,
respectively, for a {\it reference system} of overlapping
hyperspheres in dimension $d$. [See Ref. \cite{To12c} For accurate
estimates of $(\eta_c)_S$, see Ref. \cite{Qu00} for $d=2$, Refs.
\cite{Ri97b,Lo00} for $d=3$ and Ref. \cite{To12c} for $d=4--11$.]
It is noteworthy that the scaling relation
(\ref{scaling-relation}) becomes exact in the high-$d$ limit
\cite{To13a}. The aforementioned counter-intuitive relationship
between equilibrium hard-hyperparticle fluids  and the continuum
percolation model of the same overlapping hyperparticles is a
consequence of a {\it duality} relation identified in Ref.
\cite{To12a}, the fundamental and practical consequences of which
are discussed in Sec. \ref{conclusions}.

The bound (\ref{bound}) applies for a convex body
with nonzero volume ($v(K) >0$). However, it  can be generalized to apply to  an overlapping systems
of zero-volume convex  ($d-1$)-dimensional
``hyperplates" $K_H$ in $\mathbb{R}^d$ at number density $\rho$
by replacing $v(K)$ with  an appropriate ``effective volume" $v_{\mbox{\scriptsize eff} }(K_H)$
in order to define a reduced density $\eta=\rho v_{\mbox{\scriptsize eff} }(K_H)$ for hyperplates, namely,
that of a $d$-dimensional sphere of radius $r$, i.e.,
\begin{equation}
v_{\mbox{\scriptsize eff} }(K_H)= \frac{\pi^{d/2}}{\Gamma(d/2+1)}r^d,
\label{effective-vol}
\end{equation}
where $\Gamma(x)$ is the Euler-Gamma function and $r$ is a characteristic length scale of the hyperplate. Specifically,
we choose $r$ to be the radius of a spherical hyperplate that possesses the same volume of a general hyperplate $K_H$.
The scaling relation for the threshold of a hyperplate
corresponding to (\ref{scaling-relation})  was proposed to be \cite{To13a}
\begin{equation}
\eta_c
\approx \left(\frac{v_{\mbox{\scriptsize ex}}}{v_{\mbox{\scriptsize eff}}}\right)_{\mbox {\scriptsize SHP}}  \left(\frac{v_{\mbox{\scriptsize eff} }(K_H)}{v_{\mbox{\scriptsize ex}}(K_H)}\right) {(\eta_c)_{\mbox {\scriptsize SHP}}},
    \label{plate-scaling}
\end{equation}
where $(v_{\mbox{\scriptsize ex}}/v_{\mbox{\scriptsize eff}})_{\mbox {\scriptsize SHP}}$ and $(\eta_c)_{\mbox {\scriptsize SHP}}$ are, respectively, the dimensionless exclusion volume and percolation threshold of a reference system of overlapping spherical hyperplates, and $v_{\mbox{\scriptsize eff}}(K_H)$ is given by (\ref{effective-vol}).


We begin by providing  basic definitions of the exclusion volume
of a convex body in any dimension and the closely related concepts
of mixed volumes and quermassintegrals (Sec. \ref{defs}). Using
these tools from convex geometry, we then obtain a general formula
for the exclusion volume of a convex body $K$ that relates
$v_{\mbox{\scriptsize ex}}(K)$ to the corresponding
quermassintegrals  (Sec. \ref{general}). In Sec. \ref{quermass},
we obtain explicit formulas for  quermassintegrals for the variety
of specific hyperparticles described above. Then, we describe
results for the exclusion volume for oriented hyperparticles that
generally do not possess central symmetry and their corresponding
extremal values  (Sec. \ref{oriented}). Section \ref{results}
reports results for  the dimensionless exclusion volume
$v_{\mbox{\scriptsize ex}}(K)/v(K)$ for the aforementioned
specific convex-body shapes for dimensions 2 through 12. These
findings are then applied to compute the corresponding second
virial coefficients and estimates of the continuum percolation
thresholds. In Sec. \ref{conclusions}, we close with concluding
remarks, including the fundamental and practical implications of
our results.


\section{Basic Definitions and Background}
\label{defs}

We begin by defining the exclusion zones and then exclusion volume for the case of convex bodies
in $\mathbb{R}^d$ over all orientations. Then, we define the key related concepts of mixed volumes and quermassintegrals.





\subsection{Exclusion Zones}

\begin{figure}[htp]
\begin{center}
\includegraphics[height=2in,keepaspectratio,clip=]{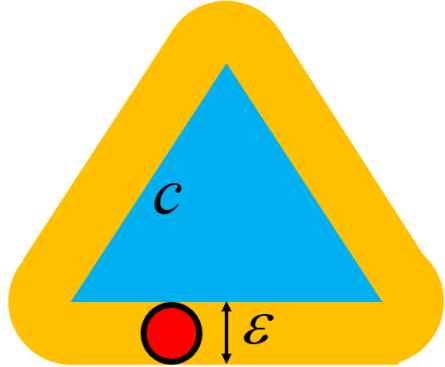}
\end{center}
\caption{(Color online) Two-dimensional illustration of the
parallel body at a distance $\varepsilon$ associated with an
equilateral triangle (two-dimensional simplex) of side length $c$,
which is the union of the blue (or dark gray) and yellow (or light
gray) regions.} \label{fig_para_body}
\end{figure}

Let $K, L \subset \mathbb{R}^d$ be convex bodies (compact nonempty convex sets). The sets $K$ and $L$ can be added via the operation of {\it Minkowski addition}:
\begin{equation}
K + L := \{x + y : x \in K, y \in L\}
\end{equation}
and a set $K$ can be multiplied by the scalar $\lambda\in\mathbb{R}$, so that
\begin{equation}
\lambda K := \{\lambda x : x \in K\}.
\end{equation}
For example, if $L=B_d$, where $B_d$ is the unit ball or sphere in $\mathbb{R}^d$,  then for $\varepsilon \ge 0$, $K + \varepsilon B_d$
 is the $\varepsilon$-neighborhood of $K$, see Fig. \ref{fig_para_body} for illustration.

With this notion of addition, we define the exclusion zone of $K$ with respect
to $L$, denoted by $Z(K, L)$, as the set of points $x$ in $\mathbb{R}^d$ such that if the centroid of $L$ were positioned at $x$, then $K$ and $L$ would intersect:
\begin{equation}
Z(K, L) := \{x \in \mathbb{R}^d : K \cap (L + x) \neq  \emptyset\}.
\end{equation}
A useful alternate expression for $Z(K, L)$, which will be more useful in calculations, is the following:
\begin{equation}
\label{exclusion-zone}
    Z(K, L) = K - L = \{x - y : x \in K, y \in L\}
\end{equation}
To show this, suppose $x \in Z(K, L)$. Then $K \cap (L + x)$ is nonempty, so that there is $y \in K$ and $z \in L$ such that $y = x + z$: it follows that $x = y - z \in K - L$, and since $x$ was arbitrary we deduce that $Z(K, L) \subset K - L$. Conversely, suppose that $x \in K - L$. Then for $y \in K$ and $z \in L$ we have $x = y - z$, so that $y = z + x \in L + x$. Thus $y \in K \cap (L + x)$ so that $K \cap (L + x) \neq \emptyset$, meaning that $x \in Z(K, L)$ and $K - L \subset Z(K, L)$. Thus $Z(K, L) = K - L$.

\subsection{Exclusion Volume Averaged Over All Rotations}

Let $\mbox{Vol}(Z(K,L))$ denote the volume of the exclusion zone $Z(K,L)$ associated with the bodies $K$ and $L$.
We define the \textit{rotationally-averaged exclusion volume} of a convex body $K$ with respect to another convex body $L$, denoted $v_{\mbox{\scriptsize ex}}(K, L)$, to be the volume of
$Z(K, L)$ averaged over all rotations of $L$ with the orientation of $K$ fixed. Let $SO(d)$ denote the set of $d$-dimensional rotations endowed
with a Haar measure $\omega$, under which all rotations are equiprobable. The rotationally-averaged exclusion volume is given by
\begin{equation}\label{full-exclusion-volume-definition}
     v_{\mbox{\scriptsize ex}}(K, L) = \int_{SO(d)}\mbox{Vol}(K + {\boldsymbol \omega}L)\,d{\boldsymbol \omega}.
\end{equation}
Notice that when $K=L$, this formula reduces to relation (\ref{v_ave}) for $v_{\mbox{\scriptsize ex}}(K)$ when the orientational probability
density function $p({\boldsymbol \omega};K)$ is set equal to unity.


\subsection{Quermassintegrals and Mixed Volumes}

In order to calculate the rotationally-averaged exclusion volume $v_{\mbox{\scriptsize ex}}(K, L)$, we need to first employ the notion of the mixed volume.
Remarkably, the volume of an $\varepsilon$-neighborhood of a convex body $K$ varies
as a polynomial in $\varepsilon$ of degree $d$ \cite{St95,Sc14,Hu20}, namely,
\begin{equation}
\mbox{Vol}(K+\varepsilon B_d) = \frac{1}{\kappa_d}\sum_{i=0}^d{d\choose i}W_i(K) \varepsilon^i
\end{equation}
where the coefficients $W_0(K),\ldots, W_d(K)$ are  called {\it quermassintegrals} (also known as Minkowski
functional or {\it intrinsic volumes} with a different normalization),
and
\begin{equation}
\kappa_d=\frac{\pi^{d/2}}{\Gamma(1+d/2)}
\label{unit-ball}
\end{equation}
 is the volume of a unit sphere.
This is the famous Steiner formula, which is a special case of the following formula:
\begin{equation}\label{mixed-volume-definition}
    \mbox{Vol}(t_1K_1 + \ldots + t_mK_m) = \sum_{i_1, \ldots, i_d = 1}^mV(K_{i_1}, \ldots, K_{i_m})t_{i_1}\ldots t_{i_m},
\end{equation}
where the  coefficients $V(K_{i_1}, \ldots, V_{i_m})$ are called \textit{mixed volumes} of the convex
bodies $K_1, \ldots, K_m \subset \mathbb{R}^d$ and $t_1, \ldots, t_m \ge 0$ are positive constants.
A particularly important fact about mixed-volumes for the problem at hand  is their behavior under averaged rotations \cite{Sc14}; specifically,
\begin{eqnarray}
\label{mixed-volume-rotation-integral}
     \int_{SO(d)}\hspace{-0.2in} && V(K_1, \ldots, K_m, {\boldsymbol \omega} K_{m + 1}, \ldots, {\boldsymbol \omega} K_d) \,d{\boldsymbol \omega} \nonumber \\
& = &\frac{1}{\kappa_d}V(K_1, \ldots,
_m, B_d[d - m])V(B_d[m], K_{m+1}, \ldots, K_d),
\end{eqnarray}
where $V(K[i], L[j])$ is a shorthand notation for  the mixed volume $V(K, \ldots, K, L, \ldots, L)$ with $i$ copies of $K$ and $j$ copies of $L$.

The Aleksandrov-Fenchel inequality, which will be used later to prove a certain bound, states
\begin{equation}
     V(K_1, K_2, K_3, \ldots, K_n)^2   \ge V(K_1, K_1, K_3, \ldots, K_n)V(K_2, K_2, K_3, \ldots, K_n).
\label{Alek}
\end{equation}
For more details, see Theorem 7.3.1 of Ref. \cite{Sc14}.

\section{General Formula for the Rotationally-Averaged Exclusion Volume}
\label{general}

\subsection{Exclusion Volume}

We now have the tools needed to derive a formula for the exclusion volume for convex bodies with random orientations.

\noindent{\bf Lemma 1.} For any convex bodies $K, L \subset \mathbb{R}^d$, the rotationally-averaged exclusion volume is explicitly
given in terms of the mixed volumes by the following formula:
\begin{equation}\label{two-body-exclusion-volume-formula}
v_{\mbox{\scriptsize ex}}(K, L) = \frac{1}{\kappa_d}\sum_{i=0}^d{d\choose i}W_{i}(L)W_{d - i}(K).
\end{equation}

\noindent{\it Proof:} From the definition of the mixed volume in relation (\ref{mixed-volume-definition}), we have
\begin{equation}\label{expansion}
    \mbox{Vol}(K + {\boldsymbol \omega} L) =  \sum_{i = 0}^d{d\choose i}V(K[i], {\boldsymbol \omega}L[d - i])
\end{equation}
Thus, equations (\ref{mixed-volume-rotation-integral}) and (\ref{expansion}) give us the following:
\begin{equation}
    \eqalign{v_{\mbox{\scriptsize ex}}(K, L) & = \int_{SO(d)} \mbox{Vol}(K + {\boldsymbol \omega}L)\,d{\boldsymbol \omega} \\ & = \int_{SO(d)}\sum_{i = 0}^d{d\choose i}V(K[i], {\boldsymbol \omega}L[d - i])\,d{\boldsymbol \omega} \\ & =
    \sum_{i=0}^d {d\choose i}\left[\frac{1}{\kappa_d}V(K[i], B_d[d-i])V(B_d[i], L[d - i])\right] \\ & = \frac{1}{\kappa_d}\sum_{i=0}^d{d\choose i}W_{d - i}(K)W_i(L).}
\end{equation}
This completes the proof.
\bigskip

\noindent{Remark:} Lemma 1 is actually a special case of the so-called {\it kinematic formula},
in which no assumption is made about the orientations of the bodies;
see, for example, Eq. (30) in Ref \cite{Me20}.

Of particular interest for us is the situation when we have two copies of the same body $K$, in which case the exclusion volume formula
of Lemma 1 simplifies as follows:
\begin{equation}\label{exclusion-volume-formula-for-one-body}
    v_{\mbox{\scriptsize ex}}(K) = v_{\mbox{\scriptsize ex}}(K, K) = \frac{1}{\kappa_d}\sum_{i=0}^d{d\choose i}W_{d - i}(K)W_i(K).
\end{equation}

\subsection{Quermassinegrals as Cross-Sectional Volumes}


Schneider \cite{Sc14} introduces the {\it area measures} $S_m(K, \omega)$ for a convex body  $K$, where $\omega$ a measurable set of rotations.
As their name suggests, they give a measure of the $m$-dimensional cross-sectional volume of $K$ averaged over $\omega$. In Ref. \cite{Sc14},
the quermassintegrals are introduced as
\begin{equation}
W_i(K) = \frac{1}{d}S_{d-i}(K, \mathbb{S}^{d-1}),
\end{equation}
which is equivalent to the definition we have given. For example,  if $K$ is a $d$-dimensional body in $\mathbb{R}^d$, then the area measure
$S_{d-1}(K, \mathbb{S}^{d-1})$ is exactly equal to the surface area of $K$, which we denote by $s(K)$.
 More precisely, $S_{d-1}(K, \mathbb{S}^{d-1})$ is equal to the $(d-1)$-dimensional Hausdorff measure of the set $\bigcup_{u \in \mathbb{S}^{d-1}}F(K, u)$, where $F(K, u)$ is the intersection of $K$ with its supporting hyperplane with normal $u$. Thus, provided that $K$ is sufficiently smooth (including polyhedra), we have
\begin{equation}
    W_1(K) = \frac{s(K)}{d}.
\label{1}
\end{equation}
Similarly, $W_{d - 1}(K)$ is related to the {\it radius of mean curvature}, ${{\bar R}(K)}(K)$, via the notion of the {\it mean width},  denoted by ${\bar w}(K)$ .
Using the fact that ${{\bar R}(K)} = {\bar w}(K)/2$ \cite{To13a}, we have that
\begin{equation}
    W_{d - 1}(K) = \frac{\kappa_d}{2}{\bar w}(K) = \kappa_d{{\bar R}(K)}.
\label{d-1}
\end{equation}
Observe that directly from the definition of the mixed volume, $V(K, \ldots, K) = v(K)$ for any convex body $K$, so that
the following two simple relationships immediately follow:
\begin{equation}
W_0(K) = v(K),
\label{0}
\end{equation}
\begin{equation}
 W_d(K) = \kappa_d.
\label{d}
\end{equation}
Other quermassintegrals in terms of  other cross-sectional volumes can be obtained from Ref. \cite{Sc14}.

\subsection{Comparison to the Torquato-Jiao Exclusion-Volume Formula}
\label{comparison}

Torquato and Jiao \cite{To13a} proposed the following formula for the exclusion volume in $d$ dimensions:
\begin{equation}\label{TJ-exclusion-volume}
 v_{\mbox{\scriptsize ex}}(K) \simeq   2 v(K) + \frac{2^d - 2}{d}s(K){{\bar R}(K)}(K),
\end{equation}
where $s(K)$ is the surface area of $K$ and ${\bar R}(K)$ is the radius of mean curvature of $K$. We show here that formula (\ref{TJ-exclusion-volume})
is exact for all bodies in dimensions 1, 2, and 3 and for greater dimensions only specific classes of bodies. Interestingly, we prove
that for $d \ge 4$ formula (\ref{TJ-exclusion-volume}) is generally  a lower bound on the exact exclusion volume, given by relation (\ref{exclusion-volume-formula-for-one-body})
for $d\ge 4$.

First, we compare the formulas (\ref{exclusion-volume-formula-for-one-body}) and (\ref{TJ-exclusion-volume}),
in the first three space dimensions.
For $d=1$, formula (\ref{exclusion-volume-formula-for-one-body}) together with (\ref{0}) and (\ref{1}) yields
\begin{equation}
    v_{\mbox{\scriptsize ex}}(K) =  W_0(K)W_1(K) = 2v(K),
\end{equation}
which agrees with formula (\ref{TJ-exclusion-volume}).
For $d=2$, formula (\ref{exclusion-volume-formula-for-one-body}) together with (\ref{0}), (\ref{d-1}) and (\ref{d}) yields
\begin{equation}
\eqalign{
    v_{\mbox{\scriptsize ex}}(K) & = \frac{1}{\kappa_2}(W_0(K)W_2(K) + 2W_1(K)^2 + W_0(K)W_2(K)) \\ & = \frac{1}{\pi}(2\pi v(K) + \frac{1}{2}s(K)^2) = 2v(K) + \frac{s(K)^2}{2\pi}
}
\end{equation}
which agrees with formula (\ref{TJ-exclusion-volume}).
For $d=3$, formula (\ref{exclusion-volume-formula-for-one-body}) together with (\ref{0}), (\ref{1}), (\ref{d-1}) and (\ref{d}) yields
\begin{equation}\eqalign{
    v_{\mbox{\scriptsize ex}}(K) & = \frac{2}{\kappa_3}(W_0(K)W_3(K) + 3W_1(K)W_2(K)) \\ & = \frac{2}{\kappa_3}\left(\kappa_3v(K) + 3\frac{s(K)}{3}( \kappa_3{\bar R}(K))\right) \\ & = 2v(K) + 2s(K){\bar R}(K).
}\end{equation}
Thus, formula (\ref{TJ-exclusion-volume}) agrees with formula (\ref{exclusion-volume-formula-for-one-body}) in dimensions 1, 2, and 3, since in these dimensions only the quermassintegrals $W_0$, $W_1$, $W_{d-1}$, and $W_d$ are relevant. In higher dimensions, however, the other quermassintegrals will generally play a role.

The Aleksandrov-Fenchel inequality (\ref{Alek}) will allow us to continue our comparison of the
formulas (\ref{exclusion-volume-formula-for-one-body}) and (\ref{TJ-exclusion-volume}).
Since the mixed volumes are symmetric \cite{Sc14}, it follows that for the quermassintegrals, we have
\begin{equation}
    W_i(K)^2 \ge W_{i-1}(K)W_{i + 1}(K).
\end{equation}
This inequality implies the inequality
\begin{equation}
    W_{i}(K)/W_{i-1}(K) \ge W_{i+1}(K)/W_i(K)
\end{equation}
and by iterating this we find that for any $j \ge i$ and sufficiently small $k$ (such that $i - k \ge 0$ and $j + k \le d$) we have
\begin{equation}
    W_i(K)W_j(K) \ge W_{i - 1}W_{j + 1}(K) \ge W_{i - k}(K)W_{j + k}(K),
\end{equation}
where the second inequality follows from iterating the first. Thus, for all $i$,
\begin{equation}
    W_i(K)W_{d-i}(K) \ge W_1(K)W_{d - 1}(K).
\end{equation}
Now we can write
\begin{equation}\label{bound-derivation}\eqalign{
    v_{\mbox{\scriptsize ex}}(K) & = \frac{1}{\kappa_d}\sum_{i=0}^d{d\choose i}W_{d - i}(K)W_i(K) \\ & = \frac{1}{\kappa_d}\left(2W_0(K)W_d(K) + \sum_{i=1}^{d-1}{d\choose i}W_i(K)W_{d-i}(K)\right) \\ & \ge \frac{1}{\kappa_d}\left(2W_0(K)W_d(K) + \sum_{i=1}^{d-1}{d\choose i}W_1(K)W_{d-1}(K)\right) \\ & = \frac{1}{\kappa_d}\left(2W_0(K)W_d(K) + (2^d - 2)W_1(K)W_{d-1}(K)\right) \\ & = \frac{1}{\kappa_d}\left(2\kappa_dv(K) + (2^d - 2)\frac{s(K)}{d}\kappa_d{\bar R}(K) \right) \\ & = 2v(K) + \frac{2^d - 2}{d}s(K){\bar R}(K).
}\end{equation}
In summary,
\begin{equation}
    v_{\mbox{\scriptsize ex}}(K) = \frac{1}{\kappa_d}\sum_{i=0}^d{d\choose i}W_{d - i}(K)W_i(K) \ge 2v(K) + \frac{2^d - 2}{d}s(K){\bar R}(K).
\label{bound-2}
\end{equation}
Thus, we have proven that the formula (\ref{TJ-exclusion-volume}) is  a rigorous lower bound on $v_{\mbox{\scriptsize ex}}(K)$.

The fact that these two formulas are linked via such an elegant argument is rather surprising, and it leaves one wondering if there is a deeper reason. The method of the proof and the fact that the formulas agree in the first three space dimensions do, nonetheless, give an intuitive explanation: both formulas are derived from an average measure of a body's cross-sectional volumes. Formula (\ref{TJ-exclusion-volume}) takes into account only one-, $(d-1)-$, and $d$-dimensional volumes, and hence does not account for information from other dimensional volumes, if they exist. Formula \ref{exclusion-volume-formula-for-one-body} does take these other contributions into account, and thus it generally produces a larger value for the exclusion volume for $d\ge 4$.

Are there situations under which  the inequality in the bound (\ref{bound-2}) becomes an equality for $d\ge 4$?
We now show that there is a class of bodies for which the two formulas do agree exactly. A sufficient condition to guarantee the bounds are equal is
that $W_i(K)^2 = W_{i-1}(K)W_{i+1}(K)$ for $i \in \{2, \ldots, d - 2\}$; then, we would have
\begin{equation}
    W_i(K)/W_{i-1}(K) = W_{i+1}(K)/W_i(K)
\end{equation}
for all $i$, and upon iterating this equality we find that
\begin{equation}
    W_i(K)W_{d - i}(K) = W_1(K)W_{d - 1}(K)
\end{equation}
for all $i$. Then all of the inequalities in relation (\ref{bound-derivation}) become equalities, and the two formulas (\ref{exclusion-volume-formula-for-one-body}) and
(\ref{TJ-exclusion-volume}) agree exactly.

It is not known in general when $W_i(K)^2 = W_{i-1}(K)W_{i+1}(K)$; however, it is known that if $K$ is a $d$-dimensional centrally symmetric convex body then $W_i(K)^2 = W_{i-1}(K)W_{i+1}(K)$ if and only if $K$ is a $(d - i - 1)$-tangential body to a sphere; see Theorem 7.6.20 of Ref. \cite{Sc14}.
If this holds for $i \in \{2, \ldots, d - 2\}$, then this is equivalent to $K$ being a 1-tangential body to a sphere, which is also called a \textit{cap body} \cite{Sc14}. A cap body is the convex hull of a sphere and a countable sequence of points $\{x_n\}$ such that for distinct $x_i$ and $x_j$, the line going through $x_i$ and $x_j$ intersects the sphere.

\section{Determination of Quermassintegrals for Specific Convex Bodies in $\mathbb{R}^d$}
\label{quermass}

With the relation (\ref{exclusion-volume-formula-for-one-body}), the problem of determining the rotationally-averaged exclusion volume
for specific convex convex bodies reduces to the problem of obtaining formulas for the corresponding quermassintegrals $W_0(K),\ldots, W_d(K)$. We obtain
such formulas in arbitrary dimension for the following convex bodies: sphere, cube,
right parallelpiped, convex cylinder, spherocylinder, general ellipsoid, ellipsoid of revolution, regular simplex,
and cross-polytope as well as lower-dimensional bodies, such as the  line segment, spherical hyperplate and cubical hyperplate. It bears repeating that the cube, regular simplex
and cross-polytope are the only regular polytopes possible for $d \ge 5$ \cite{Co73}.
For some of the convex bodies, we also provide  explicit formulas for the exclusion
volume $v_{\mbox{\scriptsize ex}}(K)$ via (\ref{exclusion-volume-formula-for-one-body}) that applies in arbitrary dimensions. In the remaining cases,
while  closed-form analytical formulas for $v_{\mbox{\scriptsize ex}}(K)$
can be presented, we do not do so because the resulting equations would be cumbersomely long.

Throughout the following discussion, $i \in \{0, \dots, d\}$ denotes the index of the quermassintegral. As the input to the quermassintegrals is obvious, we write $W_i$ instead of $W_i(K)$. We denote the surface area of the unit sphere $B_d$ in $\mathbb{R}^i$ as
\begin{equation}
\label{sphere-surface-area}
  \beta_i = d \kappa_i
\end{equation}
where
\begin{equation}
\kappa_i=\frac{\pi^{i/2}}{\Gamma(1+i/2)}
\label{sphere-volume}
\end{equation}
 is the corresponding volume of the unit sphere, as obtained from (\ref{unit-ball}).


\subsection{Spheres}
For a sphere (ball) of radius $a$, we have the following simple formula for the quermassintegrals \cite{Sa76}:
\begin{equation}
W_i = \kappa_da^{d - i}.
\end{equation}
Using the identity $\sum_{i=0}^d {d \choose i}=2^d$, formula (\ref{exclusion-volume-formula-for-one-body})
and the fact $\kappa_d$ does not depend on the index $i$, immediately leads
to the well-known result that for spheres,
\begin{equation}
\frac{v_{\mbox{\scriptsize ex}}(K)}{v(K)}=2^d.
\label{v-sphere}
\end{equation}

\subsection{Cube}

For a cube with side length $b$, we have the following simple formula \cite{Sa76}:
\begin{equation}
W_i = \kappa_i b^{d - i}
\end{equation}
Hence, according to formula
(\ref{exclusion-volume-formula-for-one-body}), we have the
explicit formula for the dimensionless exclusion volume for cubes
is given by
\begin{equation}
\frac{v_{\mbox{\scriptsize ex}}(K)}{v(K)}=\frac{1}{\kappa_d} \sum_{i=0}^d {d \choose i} \kappa_i \kappa_{d-i}.
\label{v-cube}
\end{equation}
For example, for $d=3,4,5$ and 6, we obtain from (\ref{v-cube}) the exact results $v_{\mbox{\scriptsize ex}}(K)/v(K)=11, 25.5812218\ldots, 70.75$ and
$184.3523083\ldots$, respectively.

Elementary analysis of formula (\ref{v-cube}) in the high-$d$ limit leads to the following asymptotic formula
for the dimensionless exclusion volume for the cube:
\begin{equation}
\frac{v_{\mbox{\scriptsize ex}}(K)}{v(K)} \sim \frac{2^{3(d+1)/2}}{\sqrt{3 \pi d}} \qquad (d \to +\infty).
\label{v-cube-2}
\end{equation}
Comparing formulas (\ref{v-sphere}) and (\ref{v-cube-2}), we see that the dimensionless exclusion volume for cubes relative
to that for spheres grows exponentially faster according to $2^{d/2}/\sqrt{d}$ as $d$ becomes large.
We note that this asymptotic formula already leads to very accurate predictions of $v_{\mbox{\scriptsize ex}}(K)/v(K)$ for cubes even in relatively
low  dimensions, say $d \ge 6$, as can be seen from the results presented in Sec. \ref{results}.
In addition, randomly oriented hypercubes have a much larger dimensionless exclusion volume than oriented hypercubes.

\subsection{Right Parallelepiped}

A right parallelpiped with edge lengths $b_k$, $k = \{1, \dots, d\}$ is defined to be the product $[0, b_1] \times \ldots \times [0, b_d]$. To write the formula for its quermassintegrals, let $\sigma_k$ denote the $k$th elementary symmetric polynomial on $d$ variables, namely
\begin{equation}
\sigma_k(x_1, \dots, x_d) = \sum_{1 \le i_1 \le \dots \le i_k \le d}x_{i_1}\dots x_{i_k}.
\label{eq_sigma}\end{equation}
Then we have the following expressions for $W_i(K)$ \cite{Sa76}:
\begin{equation}
W_i = \frac{\kappa_i}{{d\choose i}}\sigma_{d - i}(b_1, \dots, b_d).
\end{equation}
According to formula (\ref{exclusion-volume-formula-for-one-body}), we have that the
explicit formula for the dimensionless exclusion volume is given by
\begin{equation}
\frac{v_{\mbox{\scriptsize ex}}(K)}{v(K)}=\frac{1}{\kappa_d \Pi_{k=1}^d b_k} \sum_{i=0}^d \frac{\kappa_i \kappa_{d-i}}{{d \choose d-i}} \sigma_i \sigma_{d-i}
\end{equation}
where $\sigma_k$ is given by Eq. (\ref{eq_sigma}).

\subsection{Right Parallelepiped with a Specific Aspect Ratio}
In applications, one may often encounter right parallelpipeds $[0, b_1] \times \ldots \times [0, b_d]$ where all $b_k$ are equal ($b_k = b$) except for one, say $b_1 = h$. We call such a body a right parallelepiped with an aspect ratio $\gamma$, defined as $\gamma = h/b$. We then have the following formula for the quermassintegrals which is more efficient to calculate in simulations:

\begin{equation}W_i = \frac{\kappa_i}{{d\choose i}}\left[{d - 1 \choose d - i}b^{d - i} + \left({d \choose d - i} - {d - 1 \choose d - i}\right)\gamma b^{d - i}\right]\end{equation}

\subsection{Convex Cylinder}

A convex cylinder in $\mathbb{R}^d$ is the Cartesian product of a $(d - 1)$-dimensional sphere and an interval.
More specifically, let $B$ be a $(d -1)$-dimensional sphere of radius $a$ in $\mathbb{R}^{d - 1}$. Then let $C = (B \times \{0\}) \times [0, h]$. The body $C$ is called a convex cylinder with radius $a$ and height $h$, and we have the following formulas \cite{Sa76}:
\begin{equation}W_i = \frac{\kappa_{d-1}}{d}\left(\frac{\beta_{i-1}}{\kappa_{i-1}}a^{d - i} + (d - i)a^{d - i - 1}h\right) \quad i \ge 2,\end{equation}
and
\begin{equation}
W_0 = \kappa_{d-1}a^{d - 1}h, \quad W_1 = \frac{\kappa_{d-1}}{d}(2a^{d-1} + (d - 1)a^{d - 2}h).
\end{equation}

\subsection{Spherocylinder}

A spherocylinder is the Minkowski sum of a sphere and a line segment, and thus a spherocylinder is an $\varepsilon$-neighborhood of a line segment. Equation (13.27) of Ref. \cite{Sa76} gives a formula for the quermassintegrals of an $\varepsilon$-neighborhood of a body $K$, as follows:
\begin{equation}
W_i(K + \varepsilon B_d) = \sum_{j = 0}^{d - i}{d - i \choose j}{j}W_{i + j}(K)\varepsilon^j.
\end{equation}
When $K$ is a line segment of length $h$ and when $\varepsilon = a$, this reduces to the following formula for a spherocylinder of height $h$ and radius $a$:
\begin{equation}
W_i = a^{d-i}\kappa_d + (d - i)a^{d - i - 1}\frac{\kappa_{d-1}}{d}h.
\label{sphero}
\end{equation}
High-$d$ asymptotic analysis of the exclusion-volume formula (\ref{exclusion-volume-formula-for-one-body})
together with (\ref{sphero})  leads to the following exact scaling behavior
for the dimensionless exclusion volume of a spherocylinder with $h >0$ and finite:
\begin{equation}
\frac{v_{\mbox{\scriptsize ex}}(K)}{v(K)} \sim \frac{2^{d} \sqrt{d}}{\sqrt{32 \pi}}\, \frac{h}{a}+ \frac{2^d 3}{4}.
\label{sphero-scaling}
\end{equation}
We see that for positive, finite values of $h$, the dimensionless exclusion volume for a spherocylinder
relative to that of a sphere only rises like the square root of the dimension.

\subsection{General Ellipsoid}

An ellipsoid with axes $a_k$, $k = \{1, \dots, d\}$ is the image of a sphere with radius $1$ under the linear transformation $(x_1, \ldots, x_d) \rightarrow (a_1x_1, \ldots, a_dx_d)$.
Let $\{v_j\}_{j=1}^i$ be independent centered non-degenerate Gaussian random vectors in $\mathbb{R}^d$ whose $k$th coordinates are distributed
\begin{equation}v_j^{(k)} \sim N(0, a_k^2).\end{equation}
Let $M$ be a $i \times d$ matrix whose $j$th row is $v_j$, in other words $M = (v_1, \dots, v_i)^{T}$. Then we have the following \cite{Za14}:
\begin{equation}
W_i = \frac{\kappa_i}{{d \choose d - i}}\frac{(2\pi)^{d - i}}{(d - i)!}\mathbb{E}\left(\sqrt{\det (MM^{T})}\right),
\end{equation} where $\mathbb{E}(\cdot)$ denotes the expectation function.
While this formula cannot be calculated exactly in general,  the expectation can be readily calculated to arbitrary accuracy.

\subsection{Ellipsoid of Revolution}

An ellipsoid of revolution in $\mathbb{R}^d$ is an ellipsoid with axes $a_1, \ldots, a_d$ where every $a_k$ except for $a_1$ has the same value. Furthermore, we assume by convention that $a_1 < a_k$ for $k \neq 1$. We let the common length be denoted by $a$ and define $\lambda$ so that $a_1 = \lambda a$.

Now let $F$ denote the hypergeometric function, namely
\begin{equation}F(a, b, c; z) = \sum_{n = 0}^{\infty}\frac{(a)_n(b)_n}{(c)_n}\frac{z^n}{n!},\end{equation}
where $(x)_n$ is the Pochhammer symbol, defined as
\begin{equation}(x)_n = \frac{\Gamma(x + n)}{\Gamma(x)} = x(x + 1)\dots (x + n - 1).\end{equation}
Then we have the following Ref. \cite{Sa76}:
\begin{equation}W_i = \kappa_d\lambda^{i + 1}a^{d - i}F\left(\frac{d + 1}{2}, \frac{i}{2}, \frac{d}{2}; 1 - \lambda^2\right).\end{equation}

\subsection{Regular Simplex}

A simplex in $\mathbb{R}^d$ is the convex hull of $d + 1$ points. A regular simplex is a simplex whose edges all have equal length. For a regular simplex with edge length $c$, we have \cite{Hen97}:
\begin{equation}W_0 = \left(\frac{c}{\sqrt{2}}\right)^d\frac{\sqrt{d + 1}}{d!}, \quad\quad W_d = \kappa_d\end{equation}
For $1 \le i < d$,
\begin{equation}W_i = \left(\frac{c}{\sqrt{2}}\right)^{d - i}\frac{\kappa_i}{{d \choose d - i}}{d + 1 \choose d - i + 1}\frac{\sqrt{d - i + 1}}{d!}\gamma(T_{d-i}, T_d)\end{equation}
where $\gamma(T_{d-i}, T_d)$ is the external angle of the simplex
at a $(d - i)$-dimensional face, calculated by
\begin{equation}\gamma(T_{d - i}, T_d) = \sqrt{\frac{d - i +1}{\pi}}\int_{-\infty}^{\infty}e^{-(d - i + 1)x^2}\left(\frac{1}{\sqrt{\pi}}\int_{-\infty}^xe^{-y^2}\,dy\right)^{i}\,dx\end{equation}

\subsection{Cross-Polytope}

The canonical cross-polytope is the convex hull of the $2d$ unit vectors $e_1, -e_1, \ldots, e_d, -e_d$. We recall that a two-dimensional cross-polytope is a square, a three-dimensional cross-polytope is a regular octahedron, and a four-dimensional cross-polytope is a 16-cell. The facets
of a cross-polytope for $d\ge 3$ are regular simplices in dimension $d-1$. For a regular
cross-polytope of side length $c$ we have \cite{Hen97}:
\begin{equation}W_0 = \frac{(c\sqrt{2})^d}{d!}, \quad\quad W_d = \kappa_d\end{equation}
For $1 \le i < d$,
\begin{equation}W_i = \left(\frac{c}{\sqrt{2}}\right)^{d - i}2^{d - i + 1}\frac{\kappa_i}{{d \choose d - i}}{d \choose k + 1}\frac{\sqrt{d - i + 1}}{d!}\gamma(T_{d-i}, C_d^{\Delta})\end{equation}
where again $\gamma(T_{d-i}, C_d^{\Delta})$ denotes the external angle of the cross-polytope at a $T_{d-i}$ face, calculated by
\begin{equation}\gamma(T_{d-i}, C_d^{\Delta}) = \sqrt{\frac{k+1}{\pi}}\int_{0}^{\infty}e^{-(d - i + 1)x^2}\left(\frac{2}{\sqrt{\pi}}\int_{0}^xe^{-y^2}\,dy\right)^{i - 1}\,dx\end{equation}

\subsection{Spherical Hyperplates}

The derivation of these formulas is given in appendix \ref{appendix1}. For a spherical hyperplate of radius $a$, we have for $i = 0$ and $i = 1$
\begin{equation}
    W_0 = 0, \quad W_1 = \frac{2\kappa_{d-1}}{d}a^{d-1}
\end{equation}
and for $i \ge 2$
\begin{equation}
    W_i = \left(\frac{\kappa_{d - 1}}{\kappa_{i - 1}} \cdot \frac{i}{d}\right) \kappa_ia^{d - i}.
\end{equation}
According to Eq. (\ref{exclusion-volume-formula-for-one-body}), the
explicit formula for the exclusion volume of spherical hyperplates is given by
\begin{equation}
\frac{v_{\mbox{\scriptsize ex}}(K_H)}{v_{\mbox{eff}}(K_H)}=\frac{\Gamma(d/2+1)}{\pi^{d/2}}\left[{\frac{2 \kappa_{d-1}^3}{\kappa_d\kappa_{d-2}}\frac{(d-1)}{d^2} + \frac{1}{\kappa_d} \sum_{i=1}^{d-1} {d \choose i}\frac{\kappa_i \kappa_{d-i} \kappa^2_{d-1}}{\kappa_{i-1} \kappa_{d-i-1}} \frac{(d-i)}{d^2}}\right],
\end{equation}
where $v_{\mbox{eff}}(K_H)$ is the effective volume given  in Eq. (\ref{effective-vol}), which is the volume of a $d$-dimensional sphere with radius $r = a$. It is clear that $v_{\mbox{eff}}(K_H)$ for a spherical hyperplate must have the same high-$d$ scaling as a full-dimensional
sphere in $d$ dimensions, i.e., it must scale as $2^d$.

\subsection{Cubical Hyperplates}

As for spherical hyperplates, the derivation of the following formulas is given in \ref{appendix1}. For a cubical hyperplate of edge length $b$, we have for $i = 0$
\begin{equation}
    W_0 = 0
\end{equation}
and for $i \ge 1$
\begin{equation}
    W_i = \left(\frac{i}{d}\right) \kappa_i b^{d - i}
\end{equation}
According to Eq. (\ref{exclusion-volume-formula-for-one-body}),
the explicit formula for the exclusion volume of cubical
hyperplates is given by
\begin{equation}
\frac{v_{\mbox{\scriptsize ex}}(K_H)}{v_{\mbox{eff}}(K_H)}=\frac{\Gamma(d/2+1)}{\Gamma((d+1)/2)^{\frac{d}{d-1}}} \frac{1}{\kappa_d} \sum_{i=1}^{d-1} {d \choose i}\frac{(d-i)i}{d^2}{\kappa_{i} \kappa_{d-i}},
\label{v_cubical_plates}
\end{equation}
where the effective volume $v_{\mbox{eff}}(K_H)$ of a $d$-dimensional sphere, defined by  Eq. (\ref{effective-vol}), has radius $r=\Gamma((d+1)/2)^{1/(d-1)}b/\pi^{1/2}$. This radius corresponds to the $(d-1)$-dimensional cubical hyperplate having the  same volume as the $(d-1)$-dimensional spherical hyperplate. Similar to the case of cubes, our analysis of formula (\ref{v_cubical_plates}) in the high-$d$ limit leads to the following asymptotic formula:
\begin{equation}
\frac{v_{\mbox{\scriptsize ex}}(K_H)}{v(K_H)} \sim \frac{\sqrt{e}}{4}\frac{2^{3(d+1)/2}}{\sqrt{3 \pi d}} \qquad (d \to +\infty).
\label{v_cubical_plates2}
\end{equation}
It can be seen that the cubical hyperplates possess the same large-$d$ asymptotic scaling as cubes [c.f. Eq.(\ref{v-cube-2})], up to a constant. Comparing formulas (\ref{v-sphere}) and (\ref{v_cubical_plates2}), we see that the dimensionless exclusion volume for cubical hyperplates relative to that for spheres grows exponentially as $2^{d/2}/\sqrt{d}$ for large $d$.

\subsection{Line Segment}

Except for $W_{d-1}$ and $W_d$, all of the quermassintegrals of a line segment in $\mathbb{R}^d$ are exactly 0. This is because a line segment is intrinsically one-dimensional, and so it can only have zero- and one-dimensional cross-sectional volumes. The issue of the quermassintegrals of low-dimensional bodies is discussed in more detail and generality in \ref{appendix1}. For the line segment of length $\ell$ in $\mathbb{R}^d$, we have the following formulas \cite{Sa76}:
For $d \ge 2$,
\begin{equation}
W_i = 0 \quad (i = 0, 1, \dots, d-2), \quad W_{d - 1} = \frac{\kappa_{d-1}}{d}\ell, \quad W_d = \kappa_d.\end{equation}
For $d = 1$,
\begin{equation}W_0 = \ell, \quad W_1 = 2,
\end{equation}
which is identical to the case of overlapping rods of length
$\ell$. The relations above above together with formula
(\ref{exclusion-volume-formula-for-one-body}) prove that the
exclusion volume for a line segment in $\mathbb{R}^d$ vanishes for
$d \ge 3$.

\section{Extremal Values for Exclusion Volumes of Oriented and Non-Oriented Bodies}
\label{oriented}

\subsection{Oriented Exclusion Volume}
\label{sec-oriented}

\begin{figure}[ht]
\begin{center}
\includegraphics[height=2in,keepaspectratio,clip=]{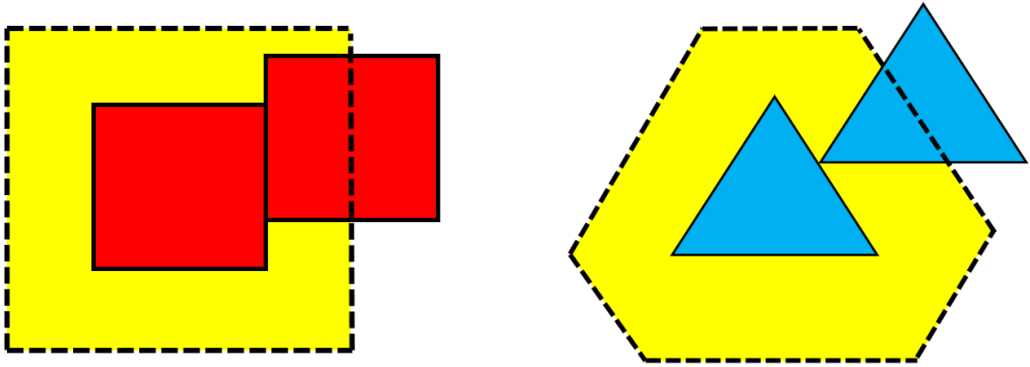}
\end{center}
\caption{(Color online) Illustration of the exclusion volume of two-dimensional  convex bodies, each with the same orientation.  In each of the two examples, the exclusion volume is the region interior to the boundary delineated  by the dashed lines.
Left panel: Centrally symmetric spherocylinder.  Right panel: Non-centrally symmetric triangle.}
\label{exclusion-2}
\end{figure}

The estimates (\ref{bound}) and (\ref{scaling-relation}) apply to systems of hyperparticles with any orientation distribution. For a system of uniformly oriented hyperparticles (Fig. \ref{exclusion-2}), there are exact results for the dimensionless quantity $v_{\mbox{\scriptsize ex}}(K)/v(K)$ of (\ref{bound}) worth discussing here.

A result known as the Roger-Shepherd inequality (Theorem 10.1.4 of Ref. \cite{Sc14}) states that for any full-dimensional convex body $K \subset \mathbb{R}^d$ (i.e., a convex body with nonzero volume), we have
\begin{equation}\label{roger}
    2^d \le \frac{v(K - K)}{v(K)} \le {2d \choose d}.
\end{equation}
The quantity $v(K - K)$ resembles the quantity $v(K - {\boldsymbol \omega} K)$ appearing in the definition of the exclusion volume, as specified
by relation (\ref{full-exclusion-volume-definition}).

If we assume that we are working in a system where all particles have an identical, fixed orientation, then it is reasonable to define an ``oriented'' exclusion volume in the same way that we defined the exclusion volume for randomly oriented particles, but without the step of averaging over orientations. Following (\ref{v_ave}), let $K$ be a convex body with some orientation ${\boldsymbol \omega}$. Then $f({\bf r}, {\boldsymbol \omega}; K)$ is the indicator function of the exclusion zone of $K$ with respect to a copy of $K$ centered at ${\bf r}$ with orientation ${\boldsymbol \omega}$. Then the \textit{oriented exclusion volume} of $K$, which we denote $v_{\mbox{\scriptsize ex}}^o$ for clarity, is given by
\begin{equation}\label{oriented-exclusion-volume}\eqalign{
    v_{\mbox{\scriptsize ex}}^{o}(K) & = \int_{\mathbb{R}^d}f({\bf r}, {\boldsymbol \omega}; K)\,d{\bf r} \\ & = \mbox{Vol}((K, K)) \\ & = v(K - K),
    }
\end{equation}
where we have used  (\ref{exclusion-zone}) and the fact that $Z(K, K) = K - K$. The randomly oriented non-spherical hyperparticles generally have a much higher dimensionless exclusion volume than the oriented ones.

\subsection{Extremal Exclusion Volumes for  Oriented Hyperparticles}

Combining the definition (\ref{oriented-exclusion-volume}) and the Roger-Shepherd inequality (\ref{roger}), we derive the following inequality for any convex, full dimensional, oriented hyperparticle $K$:
\begin{equation}\label{oriented-exclusion-volume-bounds}
    2^d \le \frac{v_{\mbox{\scriptsize ex}}^o(K)}{v(K)} \le {2d \choose d}.
\end{equation}
It is known that equality holds on the left in formula (\ref{oriented-exclusion-volume-bounds}) precisely when $K$ is centrally symmetric, and that equality holds on the right precisely when $K$ is a simplex (see Theorem 10.1.4 of Ref. \cite{Sc14}). Thus, when restricted to oriented particles, we have the fairly strong result that $v_{\mbox{\scriptsize ex}}^o(K)/v(K)$ is {\it minimized} for centrally symmetric particles and {\it maximized} for simplices, which has the high-$d$ asymptotic behavior of $2^{2d}/d^{3/2}$. This means that the ratio of $v_{\mbox{\scriptsize ex}}^o(K)/v(K)$ for simplices relative to that for spheres and other centrally symmetric bodies grows like $2^d/d^{3/2}$.

\subsection{Extremal Exclusion Volumes for  Non-Oriented Hyperparticles}
For non-oriented hyperparticles, a result of the same strength as in (\ref{oriented-exclusion-volume-bounds}) is not known for  $v_{\mbox{\scriptsize ex}}(K)/v(K)$. For randomly oriented particles, however, we can recover the lower-bound side of the inequality (\ref{oriented-exclusion-volume-bounds}).

The Brunn-Minkowski inequality (Theorem 7.1.1 of Ref. \cite{Sc14}) states that for two full dimensional convex bodes $K$ and $L$ in $\mathbb{R}^d$, we have
\begin{equation}\label{Brunn-Minkowski-Inequality}
    v(K + L)^{1/d} \ge v(K)^{1/d} + v(L)^{1/d}.
\end{equation}
Thus, for any rotation ${\boldsymbol \omega}$ we have
\begin{equation}\eqalign{
    v(K + {\boldsymbol \omega} K)^{1/d} & = v(K + {\boldsymbol \omega}K)^{1/d} \\ & \ge v(K)^{1/d} + v({\boldsymbol \omega}K)^{1/d} \\& = v(K)^{1/d} + v(K)^{1/d} \\ & = 2v(K)^{1/d}
}
\end{equation}
so that
\begin{equation}\label{brunn-minkowski-bound}
    v(K + {\boldsymbol \omega}K) \ge 2^d v(K).
\end{equation}
From the definition of the randomly oriented exclusion volume (\ref{full-exclusion-volume-definition}), we then have
\begin{equation}\eqalign{
    v_{\mbox{\scriptsize ex}}(K) = \int_{SO(d)}v(K + {\boldsymbol \omega} K)\,d{\boldsymbol \omega}  \ge \int_{SO(d)}2^dv(K)\,d{\boldsymbol \omega} = 2^dv(K)
}
\end{equation}
and we thus derive
\begin{equation}\label{randomly-oriented-extreme}
    \frac{v_{\mbox{\scriptsize ex}}(K)}{v(K)} \ge 2^d.
\end{equation}
Here, we recover the same lower bound as in the inequality (\ref{oriented-exclusion-volume-bounds}), but for randomly oriented hyperparticles.

For (\ref{oriented-exclusion-volume-bounds}), which applies to
oriented bodies, we know that the equality on the left holds
\textit{exactly} when the body is centrally symmetric. This is not
the case for randomly oriented particles, with a simple
counterexample being the cube. However, it is at least true for
spheres (balls), i.e., the lower bound is realizable by spheres.
If $B_d$ denotes the unit sphere in $\mathbb{R}^d$, then
${\boldsymbol \omega} B_d = B_d$ for any rotation ${\boldsymbol
\omega}$. Thus,
\begin{equation}
    v(B_d + {\boldsymbol \omega}B_d) = v(B_d + B_d) = 2^dv(B_d)
\end{equation}
and so
\begin{equation}\eqalign{
    v_{\mbox{\scriptsize ex}}(B_d) = \int_{SO(d)}v(B_d + {\boldsymbol \omega} B_d)\,d{\boldsymbol \omega} = \int_{SO(d)}2^dv(B_d)\,d{\boldsymbol \omega}  = 2^dv(B_d).
}
\end{equation}
We then conclude that
\begin{equation}\label{be1}
    \frac{v_{\mbox{\scriptsize ex}}(B_d)}{v(B_d)} = 2^d.
\end{equation}
In summary, from equations (\ref{randomly-oriented-extreme}) and (\ref{be1}), we have, for full dimensional randomly oriented hyperparticles, the lower bound
is realized for spheres, i.e.,
\begin{equation}
    \frac{v_{\mbox{\scriptsize ex}}(K)}{v(K)} \ge  2^d = \frac{v_{\mbox{\scriptsize ex}}(B_d)}{v(B_d)}.
\label{randomly-oriented-extreme-2}
\end{equation}

\section{Results}
\label{results}

In this section, we employ the general formula
(\ref{exclusion-volume-formula-for-one-body}) and the formulas for
the quermassintegrals presented in Sec. \ref{quermass} to
explicitly calculate the rotationally-averaged dimensionless
exclusion volume $v_{\mbox{\scriptsize ex}}(K)/v(K)$ for a variety
of selected convex bodies, including the sphere, spherocylinder,
cylinder, cube, parallelpiped, cross-polytope, simplex, as well as
spherical and cubical hyperplates in dimensions $d = 2$ through
12. To the best of our knowledge, we report exact results for the
exclusion volumes for these shapes for $d\ge 4$ for the first
time.  Our calculations for the first 6 dimensions for spheroids
indicate that they are very similar to other elongated bodies
across dimensions and the effects of elongation (i.e., aspect
ratios) are illustrated using spherocylinders. Thus, we do not
plot our results for spheroids and general ellipsoids. The
calculations for general ellipsoids are highly non-trivial for
$d\ge 4$, since their quermassintegrals involve statistical
expectations. We subsequently use the new results on
$v_{\mbox{\scriptsize ex}}(K)/v(K)$ to obtain the second virial
coefficient $B_2(K)$ for the these convex bodies in dimensions 2
through 12 using Eq. (\ref{eq_b2_vex}), as well as the estimates
of the percolation threshold $\eta_c$ across these dimensions
using scaling relation (\ref{scaling-relation}).

\subsection{Dimensionless Exclusion Volumes Across Dimensions}
\label{ex}

\begin{figure}[H]
\begin{center}
\includegraphics[height=7.5cm,keepaspectratio,clip=]{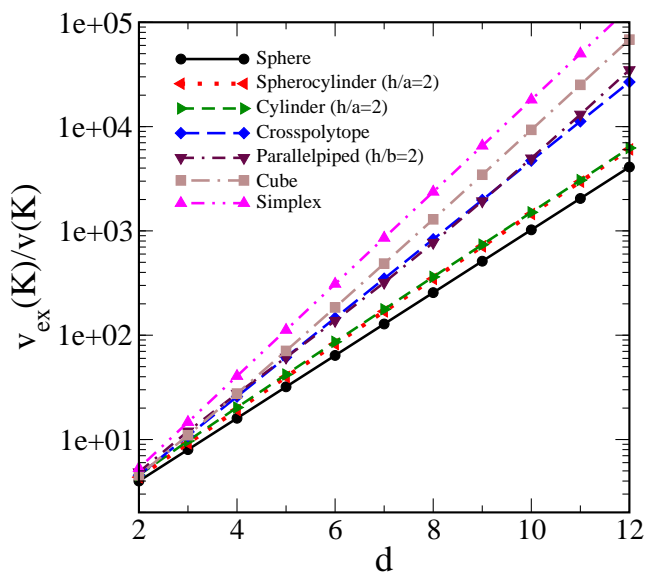}
\end{center}
\caption{(Color online) Rotationally-averaged dimensionless exclusion volume $v_{\mbox{\scriptsize ex}}(K)/v(K)$ of a convex body $K$
for selected shapes, including the sphere, spherocylinder, cylinder, cube, parallelpiped, cross-polytope, and simplex in dimensions $d = 2$ to 12. The aspect ratios of the spherocylinder, cylinder and parallelpiped are taken to be relatively small so that all of the bodies
considered are relatively compact.}
\label{fig_vex_shapes}
\end{figure}

Figure \ref{fig_vex_shapes} shows the dimensionless exclusion volume $v_{\mbox{\scriptsize ex}}(K)/v(K)$ for the sphere, cube, spherocylinder with aspect ratio $h/a=2$,
cylinder with aspect ratio $h/a=2$, parallelpiped with aspect ratio $h/b=2$, cross-polytope, and simplex in dimensions $d = 2$ to 12.
The values of $v_{\mbox{\scriptsize ex}}(K)/v(K)$ for these shapes are provided in Table \ref{tab1} in the Appendix. Our results for $d \ge 4$ appear to be new and supplement
well-known results for  these convex bodies in two and three dimensions \cite{Kih53}.  We see from the figure  that the sphere minimizes
the ratio $v_{\mbox{\scriptsize ex}}(K)/v(K)$ among the convex bodies examined. Of course, this is consistent with the rigorously exact relation
 (\ref{randomly-oriented-extreme-2})  that the sphere minimizes  $v_{\mbox{\scriptsize ex}}(K)/v(K)$ among all convex bodies.
On the other hand, simplices possess the largest ratio $v_{\mbox{\scriptsize ex}}/v$ in any dimension among the compact shapes considered here.
The cube possesses the next largest ratio $v_{\mbox{\scriptsize ex}}(K)/v(K)$ among the bodies
considered in Fig. \ref{fig_vex_shapes}. Interestingly, a least-squares fit of  the data, whether in the range $6 \le d \le 12$
or $10 \le d \le 12$ yields a robust scaling behavior. In particular, for $10 \le d \le 12$, we find $v_{\mbox{\scriptsize ex}}(K)/v(K) \sim 2^{1.44011\ldots d}$, which is close
to the exact high-$d$ asymptotic scaling (\ref{v-cube-2}), which  is controlled by the power law $2^{3d/2}$.
Similarly, a least-squares fit of  the data in Fig.  \ref{fig_vex_shapes} for simplices yields an approximate large-$d$ scaling behavior of $2^{1.6618\ldots d}$,
which will be compared below to the corresponding high-$d$ scaling behavior for oriented simplices.

Not surprisingly, the spherocylinder and cylinder with the same
aspect ratio (i.e., $h/a = 2$) possess very similar dimensionless
exclusion volumes. The spherical caps of spherocylinders lead to a
slightly smaller ratio $v_{\mbox{\scriptsize ex}}(K)/v(K)$
compared to cylinders. The cross-polytope possesses a smaller
$v_{\mbox{\scriptsize ex}}/v$ than the cube in any dimension, and
the difference increases as $d$ increases due to the former
becoming more ``isotropic" (i.e., sphere-like) in shape in higher
dimensions than cubes. The parallelpiped studied here possesses a
cubical base with edge length $b$ and height $h$, with an aspect
ratio $h/b = 2$. These parallelpipeds possess a larger ratio
$v_{\mbox{\scriptsize ex}}(K)/v(K)$ than the cube in lower
dimensions ($d=2, 3$), which then becomes smaller than that of the
cubes for $d\ge 4$. A randomly oriented parallelpiped can make
contact with another parallelpiped via either the cubical bases or
the ``rectangular'' facets. When contacting via the cubical bases,
the centroids of the particles are further separated compared to
contacts associated with the ``rectangular'' facets, which leads
to a larger exclusion volume. In higher dimensions, the number of
the ``rectangular'' facets is much larger than that of the cubical
bases. Therefore, the contribution of the ``base'' contacts to the
exclusion volume diminishes compared to the ``facet'' contacts,
leading to an orientation-averaged exclusion volume mainly
dominated by centroid separations associated with length scale $b$
(i.e., the edge length of the cubical base). On the other hand, a
larger height $h$ leads to a larger volume $v(K)$, and thus an
overall smaller dimensionless exclusion volume.

\begin{figure}[H]
\begin{center}
\includegraphics[height=7.5cm,keepaspectratio,clip=]{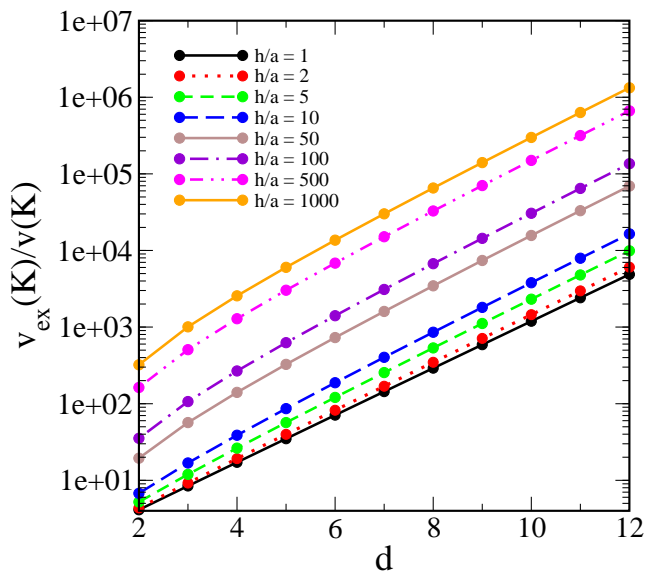}
\end{center}
\caption{(Color online) Rotationally-averaged dimensionless exclusion volume $v_{\mbox{\scriptsize ex}}(K)/v(K)$  for spherocylinders at selected aspect ratios $h/a \in [1, 1000]$ in dimensions $d = 2$ to 12. }
\label{fig_vex_sphero}
\end{figure}

To understand the effect of ``elongation" along an axis of symmetry of an anisotropic convex body with inequivalent axes on the dimensionless exclusion volume, we
 plot in Fig. \ref{fig_vex_sphero} the ratio $v_{\mbox{\scriptsize ex}}(K)/v(K)$ for spherocylinders for
selected aspect ratios $h/a$ in the interval $[1, 1000]$ in dimensions $d = 2$ to 12 and list these values in Table \ref{tab2} in the Appendix. For fixed $d$, $v_{\mbox{\scriptsize ex}}/v$
 increases significantly as the  aspect ratio $h/a$ increases, as expected.
Already for the relatively low dimensions in the range $d=6$ to $d=12$, $v_{\mbox{\scriptsize ex}}(K)/v(K)$ has a scaling behavior with $d$ that is very close
the exact high-$d$ asymptotic scaling (\ref{sphero-scaling}), i.e., it is controlled by the power law $2^d$. By comparing these results for spherocylinders
to the cases of simplices in Fig. \ref{fig_vex_shapes}, it is seen that if the aspect ratio $h/a$ is sufficiently large at fixed $d$, the ratio $v_{\mbox{\scriptsize ex}}(K)/v(K)$
for spherocylinders can exceed that for simplices. Using the exact asymptotic formula  (\ref{sphero-scaling}) and the numerically fitted scaling
of $2^{1.6618\ldots d}$ for simplices stated above, we find that, for fixed aspect ratio $h/a$, the crossover dimension $d^*$ scales like  $\ln(h/a)$, i.e.,
for $d \lesssim d^*$, $v_{\mbox{\scriptsize ex}}(K)/v(K)$ is largest for spherocylinders and for $d \gtrsim d^*$, $v_{\mbox{\scriptsize ex}}(K)/v(K)$ is largest for simplices.

\begin{figure}[H]
\begin{center}
\includegraphics[height=7.5cm,keepaspectratio,clip=]{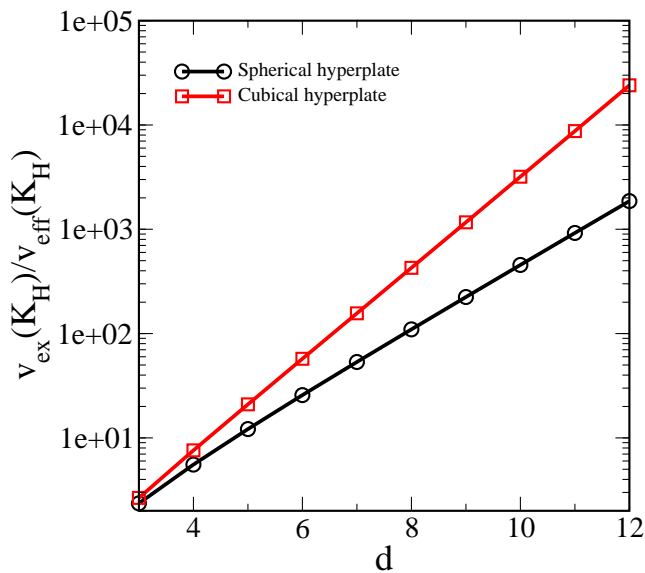}
\end{center}
\caption{(Color online) Rotationally-averaged dimensionless exclusion volume $v_{\mbox{\scriptsize ex}}(K_H)/v_{\mbox{\scriptsize eff}}(K_H)$ for spherical and cubical hyperplates in dimensions $d = 3$ to 12,
where $v_{\mbox{\scriptsize eff}}(K_H)$ is defined by (\ref{effective-vol}).}
\label{fig_vex_plates}
\end{figure}

To get a sense of the behavior $v_{\mbox{\scriptsize ex}}(K_H)/v_{\mbox{\scriptsize eff}}(K_H)$ for lower-dimensional bodies of zero volume,
we show the dimensionless exclusion volume for spherical and cubical hyperplates in dimensions $d = 3$ to 12 in Fig. \ref{fig_vex_plates}. The values of their dimensionless exclusion volumes are given in Table \ref{tab_hyperplates} in the Appendix.
It can be seen that spherical hyperplates possess a smaller value of $v_{\mbox{\scriptsize ex}}(K_H)/v_{\mbox{\scriptsize eff}}(K_H)$ than that of the cubical hyperplates, consistent with the trend for $d$-dimensional spheres and $d$-dimensional cubes. Specifically, our numerical scaling analysis of the data
in Fig. \ref{fig_vex_plates} indicates that $v_{\mbox{\scriptsize ex}}(K_H)/v_{\mbox{\scriptsize eff}}(K_H) \sim 2^{1.01446 \dots d}$ for spherical hyperplates and $v_{\mbox{\scriptsize ex}}(K_H)/v(K_H) \sim 2^{1.45687 \dots d}$ for cubical hyperplates, the latter of which is consistent with the exact asymptotic formula (\ref{v_cubical_plates2}). These results in relatively low dimensions
are consistent with the exact result   that $v_{\mbox{\scriptsize ex}}(K_H)/v_{\mbox{\scriptsize eff}}(K_H)$ for cubical hyperplates relative to that for spherical hyperplates must grow like $2^{d/2}$ for large $d$ (see Sec. \ref{quermass}). We note that it is not meaningful to compare the dimensionless exclusion volumes for zero-volume ($d-1$)-dimensional hyperplates
 to those of nonzero-volume $d$-dimensional convex bodies,
especially since the choice of the effective volume $v_{\mbox{\scriptsize eff}}(K_H)$ used to make the exclusion volume for hyperplates dimensionless is arbitrary.

\begin{figure}[H]
\begin{center}
\includegraphics[height=7.5cm,keepaspectratio,clip=]{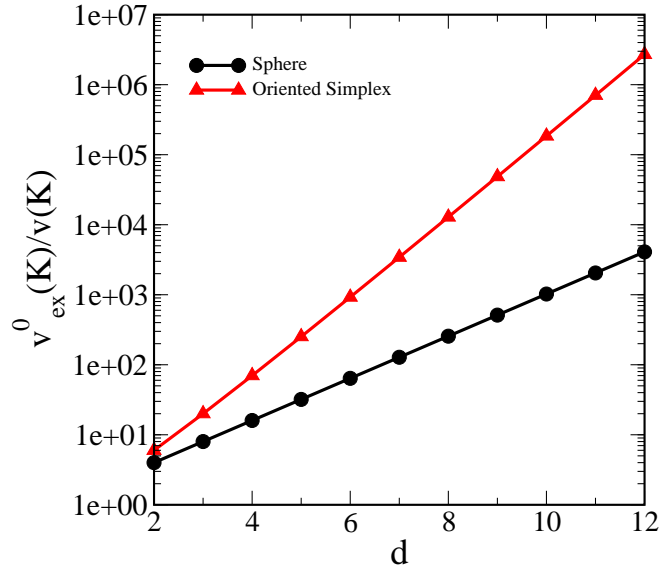}
\end{center}
\caption{(Color online) Dimensionless exclusion volume $v^0_{ex}(K)/v(K)$ for oriented simplices in dimensions $d = 2$ to 12, compared to that of spheres.}
\label{fig_vex_oriented_simplex}
\end{figure}

Finally, the dimensionless exclusion volume $v^0_{ex}(K)/v(K)$ for oriented simplices in dimensions $d = 3$ to 12 in Fig. \ref{fig_vex_oriented_simplex}
and tabulated Table \ref{tab_oriented_simplex} in the Appendix. The figure compares  results for simplices to those for spheres (or any other centrally symmetric convex body),
which rigorously achieves the minimal value of  $v^0_{ex}(K)/v(K)=2^d$ [cf. Sec. \ref{sec-oriented}]. Recall that  for oriented simplices, the large-$d$ scaling behavior of $v^0_{ex}(K)/v(K)$
is exactly given by  $2^{2d}/d^{3/2}$ [cf. Sec. \ref{sec-oriented}], which grows faster than that of spheres according to a factor of $2^{d}/d^{3/2}$ for large $d$. These
substantially different growth rates of $v^0_{ex}(K)/v(K)$ for simplices and spheres is evident in Fig. \ref{fig_vex_oriented_simplex}.  In addition,
the large-$d$ scaling for oriented simplices is exponentially larger than that of randomly oriented simplices, which we found earlier to be  $2^{1.6618\ldots d}$.

\subsection{Dimensionless Second Virial Coefficients Across Dimensions}
\label{second}

\begin{figure}[ht]
\begin{center}
\includegraphics[height=7.5cm,keepaspectratio,clip=]{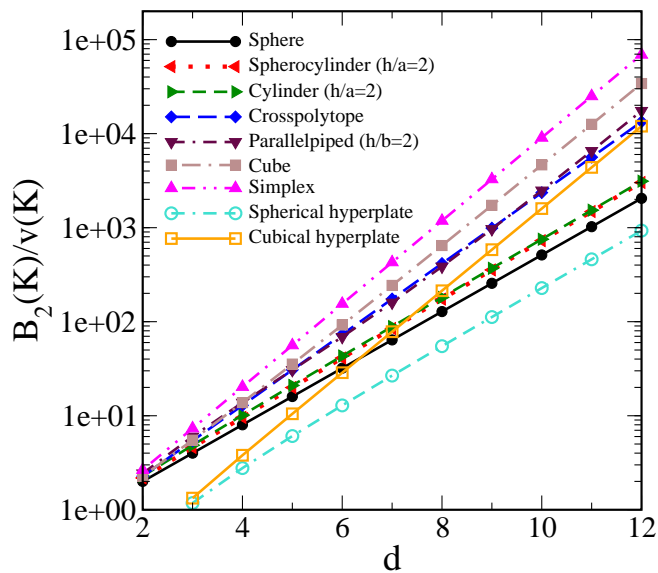}
\end{center}
\caption{(Color online) Dimensionless second virial coefficient $B_2(K)/v(K)$ for randomly oriented convex bodies for selected shapes, including the sphere, spherocylinder, cylinder, cube, parallelpiped, cross-polytope, and simplex in dimensions $d = 2$ to 12, as well as spherical and cubical hyperplates in dimensions $d = 3$ to 12.}
\label{fig_B2_shapes}
\end{figure}

\begin{figure}[ht]
\begin{center}
\includegraphics[height=7.5cm,keepaspectratio,clip=]{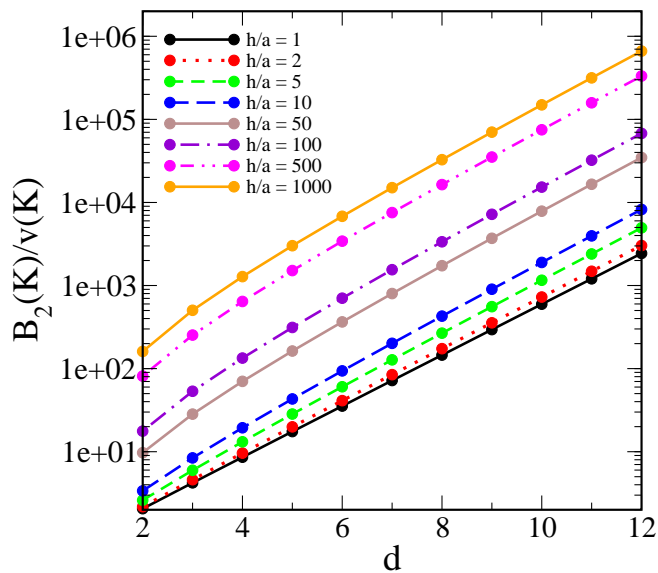}
\end{center}
\caption{(Color online) Dimensionless second virial coefficient $B_2(K)/v(K)$ for  randomly oriented spherocylinders
at selected aspect ratios $h/a \in [1, 1000]$ in dimensions $d = 2$ to 12. }
\label{fig_B2_sphero}
\end{figure}

\bigskip

Figures \ref{fig_B2_shapes} and \ref{fig_B2_sphero} shows the dimensionless second virial coefficients $B_2(K)/v(K)$ for selected convex bodies
and  spherocylinders with different aspect ratios across dimensions, respectively. Since $B_2(K)$ is trivially related to $v_{\mbox{\scriptsize ex}}(K)$ via Eq. (\ref{eq_b2_vex}), the behaviors of $B_2(K)/v(K)$ across dimensions follow exactly those for the dimensionless exclusion volume $v_{\mbox{\scriptsize ex}}(K)/v(K)$ discussed in the previous subsection. While $B_2(K)$ has long been known for
these convex bodies in two and three dimensions \cite{Kih53}, our results for $d \ge 4$ appear to be new. It is useful to reiterate that among the shapes considered, the sphere minimizes $B_2(K)/v(K)$
in any $d$ (which is rigorously true among all convex bodies) and the simplices bound $B_2(K)/v(K)$ from above, provided that the convex bodies are sufficiently compact, as shown in Fig. \ref{fig_B2_shapes}. Consistent with observations made in Sec. \ref{second},  if the aspect ratio $h/a$ is sufficiently large at fixed $d$, $B_2(K)/v(K)$
for spherocylinders can exceed that for simplices.

\subsection{Percolation Thresholds Across Dimensions}
\label{perc}

\begin{figure}[ht]
\begin{center}
\includegraphics[height=7.5cm,keepaspectratio,clip=]{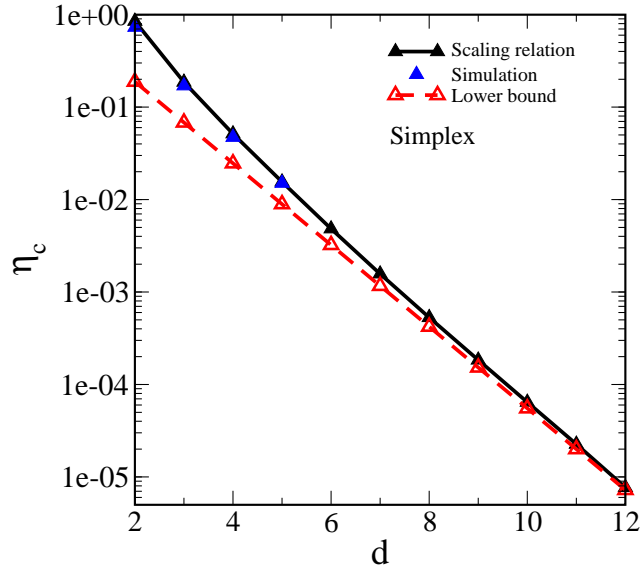}
\end{center}
\caption{(Color online) Percolation threshold $\eta_c$ for systems of randomly oriented  simplices in dimensions $d = 2$ to 12 obtained using scaling relation (\ref{scaling-relation}), which is compared to the simulation data and lower bound (\ref{bound}).}
\label{fig_eta_simplex}
\end{figure}

\begin{figure}[ht]
\begin{center}
\includegraphics[height=7.5cm,keepaspectratio,clip=]{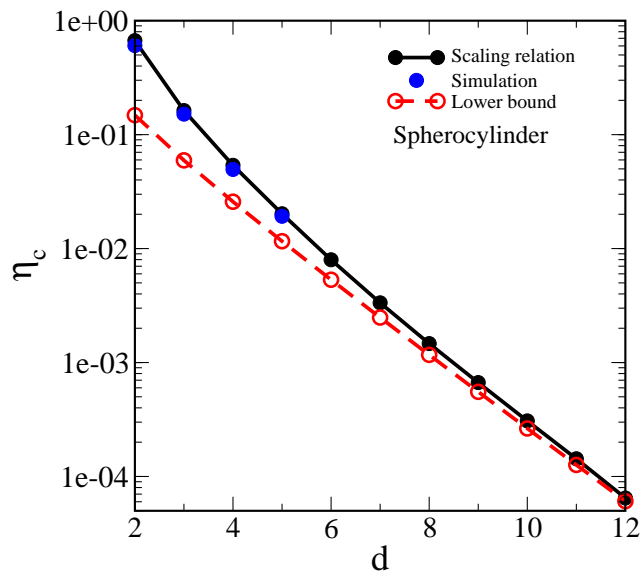}
\end{center}
\caption{(Color online) Percolation threshold $\eta_c$  for systems of randomly oriented  spherocylinders with $h/a = 10$ in dimensions $d = 2$ to 12 obtained using scaling relation (\ref{scaling-relation}), which is compared to the simulation data and lower bound (\ref{bound}).}
\label{fig_eta_sphero}
\end{figure}

In our previous work \cite{To13a}, we showed the  scaling relation (\ref{scaling-relation}), which depends on the ratio
$v_{\mbox{\scriptsize ex}}(K)/v(K)$, provides reasonably accurate estimates of the percolation threshold $\eta_c$ of many different
overlapping convex bodies in two and three dimensions when compared to corresponding simulation data. It follows from
the analysis given there that the general scaling relation (\ref{scaling-relation}) must become increasingly
accurate as $d$ becomes large for a given convex body. This is yet another manifestation
of the principle that low-dimensional percolation properties
encode high-dimensional information \cite{To12a}. In light of the fact that the formula for dimensionless
exclusion volume $v_{\mbox{\scriptsize ex}}(K)/v(K)$ given in Ref. \cite{To13a} of a convex body is generally a lower bound
on this quantity for $d\ge 4$ (see Sec. \ref{comparison}), this means that the scaling estimates for $\eta_c$
for the selected convex shapes  given there were generally overestimated for $d \ge 4$.
Thus, our interest here is in evaluating the accurate scaling relation  (\ref{scaling-relation})
across dimensions using the exact expressions for  $v_{\mbox{\scriptsize ex}}(K)/v(K)$ for the aforementioned convex bodies
given in the present paper.

While we expect the scaling relation (\ref{scaling-relation}) to be already  very accurate
for $d=4$ and greater dimensions, we confirm this expectation by carrying out
computer simulations of the percolation threshold $\eta_c$ for spherocylinders and regular simplices
for dimensions 2 through 5 using the rescaled-particle simulation method discussed in detail in Ref.~\cite{To12c}.
Spherocylinders are centrally symmetric bodies and allow us to investigate
the effects of elongation on the  theoretical estimates of $\eta_c$ based on the scaling relation across
dimensions, which is a challenging case to predict. We chose to simulate percolation of simplices
because they are compact, noncentrally symmetric bodies.


Figure \ref{fig_eta_simplex} compares the rescaled-particle simulation results  for simplices to the scaling relation (\ref{scaling-relation}) as well as to the
 lower bound (\ref{bound}). Figure \ref{fig_eta_sphero} shows the corresponding plot for spherocylinders.
A crucial observation to be made from the figures is how closely
the scaling relation (\ref{scaling-relation}) predicts the
simulated values of $\eta_c$ for the both simplices and
spherocylinders across the relatively low  dimensions from $d=2$
through $d=5$. For reasons noted earlier, the scaling relation
will yield analytical predictions with increasing accuracy as $d$
increases and becomes exact in $d\rightarrow\infty$. This can also
be seen from the convergence of the scaling relation prediction
and the rigorous lower bound, the latter of which becomes exact in
the high-$d$ limit \cite{To13a}. Using the high-$d$ scalings
reported in Sec. \ref{ex} together with scaling relation
(\ref{scaling-relation}) enables us to conclude that the decay of
$\eta_c$ with $d$ is controlled by the inverse power law
$2^{-1.6618d}$ for simplices and   the inverse power law $2^{-d}$
for spherocylinders.

\begin{figure}[ht]
\begin{center}
\includegraphics[height=7.5cm,keepaspectratio,clip=]{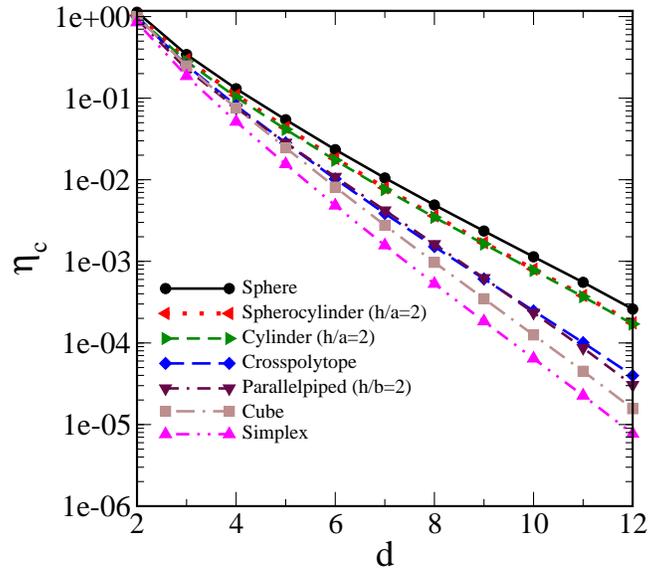}
\end{center}
\caption{(Color online) Percolation threshold $\eta_c$ for systems of randomly oriented convex bodies for selected shapes, including the sphere, spherocylinder, cylinder, cube, parallelpiped, cross-polytope, and simplex in dimensions $d = 2$ to 12, as obtained from scaling relation (\ref{scaling-relation}).}.
\label{fig_eta_shapes}
\end{figure}

Having further verified the accuracy of the scaling relation
for simplices and spherocylinders, we now employ it to predict $\eta_c$ for other shapes. Figure \ref{fig_eta_shapes} shows such estimates of $\eta_c$
for selected convex bodies across dimensions. Since $\eta_c$ is inversely proportional to the dimensionless exclusion volume  $v_{\mbox{\scriptsize ex}}(K)/v(K)$ [c.f. Eq. (\ref{scaling-relation})],
the behavior of $\eta_c$ for different shapes at fixed $d$ is the opposite of the trends described in Sec. \ref{ex} for the corresponding $v_{\mbox{\scriptsize ex}}(K)/v(K)$.
For example, among the compact shapes with finite volumes, spheres possess the largest threshold $\eta_c$, and simplices possess the smallest value of $\eta_c$, whether they are randomly oriented or uniformly oriented. In fact,
according to the rigorous relations (\ref{oriented-exclusion-volume-bounds}) and (\ref{randomly-oriented-extreme-2}) and Ref. \cite{To12a}, the sphere provably possesses
the maximal threshold among all such nonzero-volume convex bodies in the high-$d$ limit. Thus, we
conjecture that overlapping spheres possess the maximal value of $\eta_c$ among all
identical nonzero-volume convex overlapping bodies,  randomly or uniformly oriented, for $d \ge 2$. Furthermore, in light of the upper bound (\ref{oriented-exclusion-volume-bounds}),
we conjecture that among all oriented nonzero-volume convex bodies, overlapping simplices have the minimal value of $\eta_c$ for $d\ge 2$.

\begin{figure}[ht]
\begin{center}
\includegraphics[height=7.5cm,keepaspectratio,clip=]{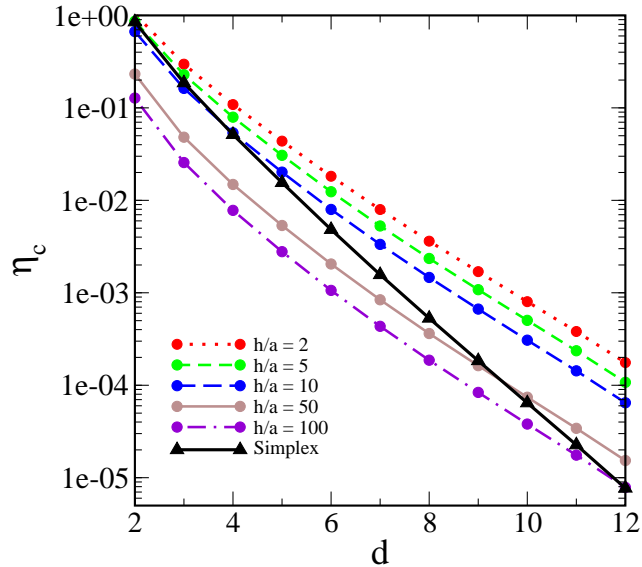}
\end{center}
\caption{(Color online) Comparison of the percolation threshold $\eta_c$ for randomly oriented spherocylinders with various aspect ratios $h/a$
to that of randomly oriented simplices across dimensions, as obtained using the scaling relation (\ref{scaling-relation}).}
\label{fig_eta_simplex_sphero}
\end{figure}

While randomly oriented simplices yield the lowest percolation thresholds among
the convex bodies considered, provided that they are relatively compact, elongated shapes, such as spherocylinders,
can have a lower threshold if their aspect ratio $h/a$ is sufficiently large. These distinctions
between the percolation thresholds of these two convex bodies are clearly seen in
Fig. \ref{fig_eta_simplex_sphero}. Consistent with the results reported in Sec. \ref{ex}, we see
that for a fix aspect ratio $h/a$, there is a  crossover dimension $d^* \sim \ln(h/a) $ beyond which the simplices possess
smaller values of  $\eta_c$ compared to that of spherocylinders. We also see from Fig. \ref{fig_eta_shapes} that cubes possess a smaller value of $\eta_c$ than
that of cross-polytopes; and cylinders possess a smaller value of $\eta_c$ than that of spherocylinders with the same aspect ratio. In the case of cubes, exact high-$d$ asymptotic formula (\ref{v-cube-2}) reported  together with scaling relation (\ref{scaling-relation})
enables us to conclude that the decay of $\eta_c$ with $d$ is controlled by the inverse power law  $2^{-3d/2}$.

\begin{figure}[ht]
\begin{center}
\includegraphics[height=7.5cm,keepaspectratio,clip=]{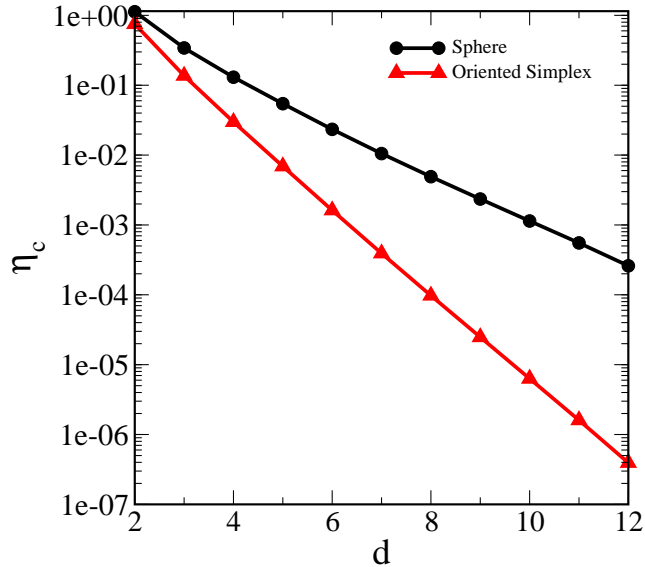}
\end{center}
\caption{(Color online) Percolation threshold $\eta_c$  for uniformly oriented simplices and spheres in dimensions $d = 2$ to 12, as obtained using the scaling relation (\ref{scaling-relation}).}
\label{fig_eta_oriented_simplex}
\end{figure}

\begin{figure}[ht]
\begin{center}
\includegraphics[height=7.5cm,keepaspectratio,clip=]{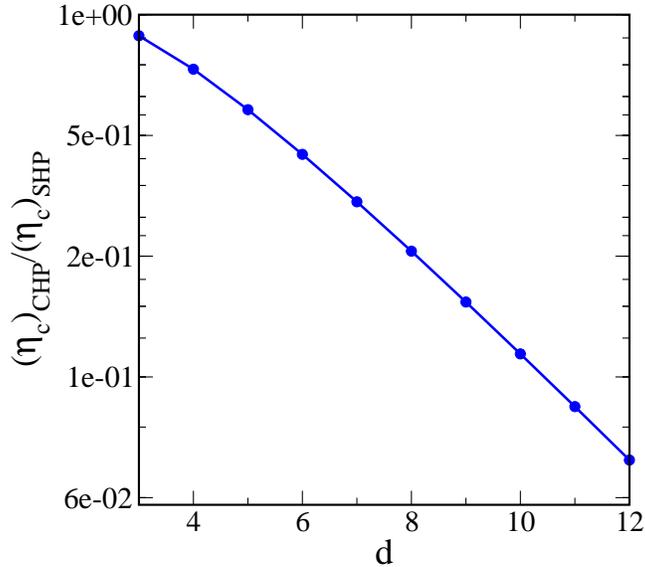}
\end{center}
\caption{(Color online) Ratio between the percolation threshold of cubical hyperplates $(\eta_c)_{\mbox{\scriptsize CHP}}$ and that of spherical hyperplates $(\eta_c)_{\mbox {\scriptsize SHP}}$ dimensions $d = 3$ to 12 obtained using the scaling relation (\ref{plate-scaling}).}
\label{fig_eta_hyperplate}
\end{figure}


Figure \ref{fig_eta_hyperplate} shows the ratio between the percolation threshold of cubical hyperplates $(\eta_c)_{\mbox {\scriptsize CHP}}$ and that of spherical hyperplates $(\eta_c)_{\mbox {\scriptsize SHP}}$ in dimensions $d = 3$ to 12 obtained using the scaling relation (\ref{plate-scaling}). In principle, the scaling relation (\ref{plate-scaling}) allows one to obtain accurate estimates of the percolation threshold for nonspherical hyperplates, given  accurate values of $(\eta_c)_{\mbox {\scriptsize SHP}}$ for the reference
spherical hyperplate system. However, such values are not available,  except for $d=3$ \cite{Yi09}. In Ref.~\cite{To13a},
we showed that Eq. (\ref{plate-scaling}) indeed led to very accurate estimates of the percolation thresholds for various two-dimensional plates in three-dimensional space, including square,
triangular, elliptical and rectangular  plates. Such good agreement already for $d=3$ means that the scaling relation
(\ref{plate-scaling}) should become increasingly more accurate as $d$ increases above three.
Figure \ref{fig_eta_hyperplate} shows  the percolation threshold of cubical hyperplates relative to that of spherical
hyperplates descends exponentially fast with $d$, namely, it decays like   $2^{-d/2}$, which is obtained
using the high-$d$ scalings reported in Sec. \ref{quermass} together with scaling relation (\ref{plate-scaling}).
In analogy with the conjectures made above for full $d$-dimensional bodies, we conjecture that among all the convex hyperplates, spherical
hyperplates have the largest percolation thresholds for any fixed $d\ge 3$. We emphasize again that in general, results for $(d-1)$-dimensional
bodies should not be compared to those for full $d$-dimensional bodies, especially because of
the arbitrary choice used for the effective volume $v_{\mbox{\scriptsize eff}}(K_H)$ of a zero-volume hyperplate
in (\ref{plate-scaling}) to make its exclusion volume dimensionless, as stressed in Sec.~\ref{ex}.


Finally, we note that randomly oriented non-spherical hyperparticles generally have a much smaller threshold than that of their oriented counterparts due to the theorems presented in Sec. \ref{oriented}. This can be seen by comparing the curves of  $\eta_c$ of oriented simplices, which must decay
like $2^{-2d}$, shown in Fig. \ref{fig_eta_oriented_simplex}
to that of $\eta_c$ of randomly oriented simplices, which decays
like $2^{-1.6618}$, shown in Fig. \ref{fig_eta_shapes}. Figure \ref{fig_eta_oriented_simplex} also includes the estimate
of $\eta_c$ for spheres.


\bigskip

\section{Conclusions}
\label{conclusions}

In this paper, we have provided a general formula for the exclusion volume $v_{\mbox{\scriptsize ex}}(K)$
for an arbitrary convex body $K$ in any space dimension, including both the rotationally-averaged exclusion volume and the exclusion volume associated with uniform
orientations of $K$. We showed that the sphere minimizes the dimensionless exclusion volume $v_{\mbox{\scriptsize ex}}(K)/v(K)$
among all convex bodies, whether randomly oriented or uniformly oriented, for any $d$.
When the bodies have the same orientation, the simplex maximizes the dimensionless exclusion volume $v_{\mbox{\scriptsize ex}}(K)/v(K)$ for any $d$
with a large-$d$ asymptotic scaling behavior of $2^{2d}/d^{3/2}$.
 We demonstrated that the rotationally-averaged exclusion volume $v_{\mbox{\scriptsize ex}}(K)$ can be written as certain weighted sums of quermassintegrals $W_0(K),\ldots, W_d(K)$ of $K$.
Subsequently, we presented explicit expressions for quermassintegrals for various nonspherical convex bodies, including cubes, parallelepipeds, regular simplices, cross-polytopes,
cylinders, spherocylinders, ellipsoids as well as lower-dimensional bodies, such as hyperplates and line segments. For certain shapes, explicit formula and large-$d$ asymptotic expressions of $v_{\mbox{\scriptsize ex}}(K)$ are obtained. These results were used to evaluate the  rotationally-averaged  ratio $v_{\mbox{\scriptsize ex}}(K)$ for these convex-body shapes for dimensions 2 through 12. While the sphere is the minimal shape, we showed that among the convex bodies considered that are sufficiently compact, the simplex possesses the maximal $v_{\mbox{\scriptsize ex}}(K)/v(K)$ with a scaling behavior of $2^{1.6618\ldots d}$, which grows more slowly than the corresponding ratio
for oriented simplices.

The exclusion volume results were subsequently utilized to
determine the corresponding second virial coefficient
$B_2(K)/v(K)$ of the hard hyperparticles that we considered  for
the first time. Such information allows us to draw some
conclusions on the effect of body shape on  the disorder-order
equilibrium phase transition in relatively low dimensions for
several reasons. First, we have demonstrated that the scaling
behavior of $v_{\mbox{\scriptsize ex}}(K)/v(K)$ or, equivalently,
$B_2(K)/v(K)$ for a range of relatively low dimensions considered
(from $d=6$ to $d=12$) agrees well with the exact high-$d$
asymptotic scalings. This further supports the general principle
that high-dimensional information is encoded in relatively low
dimensions \cite{To12a,To12c,To13a}. Second, we noted earlier that
the  dominant contribution to the pressure of a hard-hyperparticle
equilibrium fluid is given by the truncation of the virial
expansion through second-order terms [cf. (\ref{virial})]  in the
high-$d$ asymptotic limit \cite{Fr99}. Thus, in sufficiently high
dimensions, we expect that hyperparticles with a larger
dimensionless exclusion volume $v_{\mbox{\scriptsize ex}}(K)/v(K)$
should have an entropy-driven disorder-order transition occurring
at a lower density, since hyperparticles with larger exclusion
volumes impose non-trivial correlations among their neighbors at
much lower densities than those with smaller exclusion volumes.
The same idea was used by Onsager to discover a nematic phase
transition for needle-like particles in three dimensions
\cite{On49}.

We also applied our results to compute estimates of the continuum
percolation threshold $\eta_c$ using a scaling relation derived
previously by the authors for systems of identical overlapping
convex bodies. It is noteworthy that while the scaling relation
becomes exact in $d\rightarrow \infty$, it already yields very
accurate predictions even in relatively low dimensions. The
accuracy of the scaling relation predictions is ascertained using
numerical simulations for simplices and spherocylinders in
dimensions 2 through 5, which verify that these estimates indeed
become increasingly accurate as the space dimension increases.
Among the shapes with nonzero volume that we examined, we showed
that spheres possess the largest threshold $\eta_c$, and simplices
possess the smallest value of $\eta_c$, whether they are randomly
oriented or uniformly oriented. We conjectured that overlapping
spheres, possess the maximal value of $\eta_c$ among all identical
nonzero-volume convex overlapping bodies , randomly or uniformly
oriented, for $d \ge 2$. We also conjectured that, among all
identical, oriented nonzero-volume convex bodies, overlapping
simplices have the minimal value of $\eta_c$ for $d\ge 2$.
Similarly, we conjecture that among all the convex hyperplates,
spherical hyperplates have the largest percolation thresholds for
any fixed $d\ge 3$.

It should not go unnoticed that the scaling relations for $\eta_c$
that utilize the exact general explicit expressions for the
exclusion volume readily allow one to estimate the percolation
threshold of a wide spectrum of hyperparticles across dimensions,
well beyond the specific choices of shape parameters and
dimensions that we explicitly studied here. Importantly, our
numerical results indicate that the estimates of $\eta_c$ are
already reasonably accurate in three dimensions, which opens up
many practical applications of our results in physics and material
science problems that account for the effect of particle shapes.

In the Introduction, we noted the duality relation between the
continuum percolation of overlapping hyperparticles and the
equilibrium hard-hyperparticle fluids of the same shape
\cite{To12a}. Combination of this duality relation with the
so-called decorrelation principle for disordered
hard-hyperparticle packings \cite{ToSt06, ZaTo11}, implies that in
sufficiently high dimensions the percolation threshold $\eta_c$ of
overlapping hyperparticles is directly related to the
disorder-order phase transition density (i.e., the freezing-point)
of the corresponding equilibrium hard-hyperparticle fluid
\cite{To12a}. This is an outstanding open problem for our future
research.


Finally, we note that our results for the dimensionless oriented
exclusion volume $v^o_{\mbox{\scriptsize ex}}(K)/v(K)$ of convex
body $K$ has implications for the optimal packing of $K$
\cite{To09b}. Specifically, it has been conjectured \cite{ToJi12}
that the optimal packing of a centrally symmetric convex body with
equivalent principal axes (e.g., an octahedron) is achieved by the
associated optimal Bravais-lattice packing in which all the bodies
are aligned; while the optimal packing of a body without central
symmetry (e.g., a tetrahedron) is generally given by a
non-Bravais-lattice packing in which the bodies have different
orientations. Our current study further supports these organizing
principles, i.e., $v^o_{\mbox{\scriptsize ex}}(K)/v(K)$ is
minimized when the bodies are aligned for a centrally symmetric
body, while nonaligned orientations can result in a much smaller
$v^o_{\mbox{\scriptsize ex}}(K)/v(K)$ for noncentrally symmetric
shapes. Moreover, our new results on the exclusion volumes for a
wide spectrum of convex bodies across dimensions suggest that
similar organizing principles for the densest packings could also
hold in higher dimensions ($d \ge 4$), which we will explore in
our future studies.


\section*{Acknowledgement}
We are deeply grateful to Yair Shenfeld who made us aware of the relationship
of  quermassintegrals to exclusion volumes and how our previous expression
for the latter is a lower bound. We thank Alexander McWeeney for his assistance
with preliminary calculations using the rescaled particle method. This work was supported  by the National Science Foundation under Grant No. CBET-1701843.

\appendix

\section{Quermassintegrals of Lower-Dimensional Bodies and Hyperplates}
\label{appendix1}

As discussed in the Introduction, it is also of interest to consider lower-dimensional bodies and hyperplates in $\mathbb{R}^d$. These bodies
in $\mathbb{R}^d$ have zero volume, and so an effective volume (\ref{effective-vol}) is used in the bound (\ref{bound}). The exclusion volume of a hyperplate, however, is defined exactly as in Sec. \ref{defs}, and the formula (\ref{exclusion-volume-formula-for-one-body}) still applies. Thus to calculate the exclusion volume of a hyperplate, we must be able to calculate its quermassintegrals. It turns out that there is a very general way to find the quermassintegrals of a low-dimensional body in $\mathbb{R}^d$. By a low-dimensional body,
we mean a body whose intrinsic dimension is $d - 1$ or less.

\subsection{Intrinsic Volumes}

The quermassintegrals of a body $K \subset \mathbb{R}^d$ measure the cross-sectional volumes of $K$ in $\mathbb{R}^d$, which was explained in some detail in section \ref{comparison}. Intuitively, the cross-sectional volumes of a body $K$ should be the same regardless of the dimension of the ambient space $\mathbb{R}^d$ in which $K$ is embedded. However, the quermassintegrals of a body do actually depend on the dimension of the embedding space: nevertheless, there is a normalization of the quermassintegrals, called the \textit{intrinsic volumes}
or Minkowski functionals, which is invariant with respect to the dimension of the embedding space.

To be more precise, fix a body $K$ and an embedding space $\mathbb{R}^d$. We have already defined the quermassintegrals $W_0(K), \ldots, W_d(K)$. We now define a new set of functions, $V_0(K), \ldots, V_d(K)$, called the \textit{intrinsic volumes} of $K$, in the following manner \cite{Sc14}:
\begin{equation}\label{Intrinsic-Volume-Definition}
    V_i(K) = \frac{1}{\kappa_{d - i}}{d \choose i}W_{d - i}(K),
\end{equation}
where $\kappa_{d - i}$ is the volume of the unit sphere in $\mathbb{R}^{d - i}$, defined in relation (\ref{sphere-volume}); we take the additional convention that $\kappa_{0} = 1$. This definition makes $V_i$ a measure of the $i$-dimensional cross-sectional volume of $K$.

As it turns out, the definition we have given for the functions $V_i$ does not depend on the definition of the ambient space. That is, suppose $K$ is a body in $\mathbb{R}^d$, with intrinsic volumes $V_0(K), \ldots, V_d(K)$ with respect to $\mathbb{R}^d$. Now place $K$ into the space $\mathbb{R}^{D}$ with $D > d$, and let $V'_0(K), \ldots V_d'(K), \ldots, V_D'(K)$ be the intrinsic volumes of $K$ with respect to $\mathbb{R}^D$. Then, for $0 \le i \le d$, we have $V_i(K) = V_i'(K)$.

Using this fact, we can find the quermassintegrals of a body in a lower dimensional space to calculate its quermassintegrals in a higher dimensional space. Let $K$ be a convex body in $\mathbb{R}^d$, with quermassintegrals $W_0', \ldots, W_d'$ and intrinsic volumes $V_0, \ldots, V_d$. We want to calculate the quermassintegrals $W_0, \ldots, W_D$ for $K$ embedded in $\mathbb{R}^D$ for $D > d$.

To do this, we first note the following: if $K$ can be embedded in
$\mathbb{R}^d$, then $K$ is at most $d$-dimensional. In
$\mathbb{R}^D$, this means that $V_i(K) = 0$ for $d < i \le D$.
This follows formally from Eq. (4.23) of Ref. \cite{Sc14};
intuitively, this follows from the idea that if $K$ is at most
$d$-dimensional, then its cross-sectional volumes of dimension
greater than $d$ should be 0.

We now have the following from relation (\ref{Intrinsic-Volume-Definition}):
\begin{equation}\eqalign{
    W_i(K) = \kappa_i {D \choose D - i}^{-1}V_{D-i}(K)
}\end{equation}
Thus, if $D - i > d$, then $W_i(K) = 0$. If $D - i \le d$, then
\begin{equation}\eqalign{
    W_i(K) & = \kappa_i {D \choose D - i}^{-1}V_{D-i}(K)  \\ & = \kappa_i {D \choose D - i}^{-1}\left(\frac{1}{\kappa_{d - D + i}}{d \choose d - D + i}W'_{d - D + i}(K)\right) \\ & = \frac{\kappa_i}{\kappa_{d - D + i}}{D \choose D - i}^{-1}{d \choose d - D + i}W'_{d - D + i}(K).
}\end{equation}

In summary, we have
\begin{equation}\label{higher-dimension-quermassintegral-formula}
    W_i(K) = \left\{
    \begin{array}{ll}
      \frac{\kappa_i}{\kappa_{d - D + i}}{D \choose D - i}^{-1}{d \choose d - D + i}W'_{d - D + i}(K) & D - i \le d \\
      \quad & \\
      0 & D - i > d
    \end{array} \right.
\end{equation}

\subsection{Application to Spherical Hyperplates}


Here, we apply formula (\ref{higher-dimension-quermassintegral-formula}) to spherical hyperplates. A spherical hyperplate is a $(d - 1)$-dimensional sphere in $\mathbb{R}^d$. Suppose we have a spherical hyperplate $H$ of radius $a$. Then, in $\mathbb{R}^{d-1}$, we have
\begin{equation}
    W_i'(H) = \kappa_{d - 1}a^{d - 1 - i}.
\end{equation}
Thus, in $\mathbb{R}^d$, we have $W_0(H) = 0$, and for $i \ge 1$
\begin{equation}\eqalign{
    W_i(H) & = \frac{\kappa_i}{\kappa_{i - 1}}{d \choose d - i}^{-1}{d - 1 \choose i - 1}W'_{i-1}(H) \\ & = \left(\frac{\kappa_{d - 1}}{\kappa_{i - 1}} \cdot \frac{i}{d}\right) \kappa_i a^{d - i}
}\end{equation}
It is interesting to observe that if $C$ is a cube with edge length $a$, then the above formula says that
\begin{equation}
    W_i(H) = \left(\frac{\kappa_{d - 1}}{\kappa_{i - 1}} \cdot \frac{i}{d}\right)W_i(C).
\end{equation}

\subsection{Application to cubical hyperplates}
We now apply (\ref{higher-dimension-quermassintegral-formula}) to cubical hyperplates. A cubical hyperplate is a $(d - 1)$-dimensional cube in $\mathbb{R}^d$. Suppose we have a cubical hyperplate $H$ with edge length $b$. Then, in $\mathbb{R}^{d-1}$, we have
\begin{equation}
    W_i'(H) = \kappa_{i}b^{d - 1 - i}.
\end{equation}
Thus, in $\mathbb{R}^d$, we have $W_0(H) = 0$, and for $i \ge 1$
\begin{equation}\eqalign{
    W_i(H) & = \frac{\kappa_i}{\kappa_{i - 1}}{d \choose d - i}^{-1}{d - 1 \choose i - 1}W'_{i-1}(H) \\ & = \left(\frac{i}{d}\right) \kappa_i b^{d-i}
}\end{equation}
Once again, we observe that if $C$ is a cube of edge length $b$, then the above formula says that
\begin{equation}
    W_i(H) = \left(\frac{i}{d}\right)W_i(C).
\end{equation}
Furthermore, if $H_S$ is a spherical hyperplate of radius $a$, then
\begin{equation}
    W_i(H) = \left(\frac{\kappa_{d-1}}{\kappa_{i-1}}\right)^{-1}W_i(H_S).
\end{equation}

\subsection{Dimensionless Exclusion Volume for Selected Shapes Across Dimensions }

In the main paper, we graphically show the dimensionless exclusion
volume $v_{\mbox{\scriptsize ex}}(K)/v(K)$ for selected shapes in
dimensions 2 through 12. Here, we provide the values for these
shapes, which are provided in Tables A1 to A4.

\begin{table}[htp]
\centering
\begin{tabular}{l|@{\hspace{0.3cm}}c|@{\hspace{0.3cm}}c|@{\hspace{0.3cm}}c|@{\hspace{0.3cm}}c|@{\hspace{0.3cm}}c|@{\hspace{0.3cm}}c}
\hline\hline
& Sphere & Cylinder & Cross-polytope & Parallelpiped & Cube & Simplex  \\
\hline
$d$=2  & 4 & 4.54648 & 4.54648 & 4.86479 & 4.54648 & 5.30797  \\\hline
$d$=3  & 8 & 9.71239 & 10.8301 & 12 & 11  & 14.6726   \\\hline
$d$=4  & 16 & 20.3032 & 25.7981 & 27.5211 & 27.5812  & 40.5589    \\\hline
$d$=5  & 32 & 42.0105 & 61.453 & 61.25 & 70.75  & 112.115   \\\hline
$d$=6  & 64 & 86.4313 & 146.386 & 137.647 & 184.352  & 309.916   \\\hline
$d$=7  & 128 & 177.184 & 348.702 & 319.417 & 485.875  & 856.69    \\\hline
$d$=8  & 256 & 362.325 & 830.635 & 770.823 & 1291.69  & 2368.11   \\\hline
$d$=9  & 512 & 739.563 & 1978.64 & 1928.83 & 3457.3  & 6546.0    \\\hline
$d$=10 & 1024 & 1507.39 & 4713.27 & 4968.09 & 9304.28  & 18095.1   \\\hline
$d$=11 & 2048 & 3068.79 & 11227.4 & 13069.1 & 25152  & 50019.5    \\\hline
$d$=12 & 4096 & 6241.36 & 26744.5 & 34886.3 & 68247.7  & 138267   \\\hline
\hline
\end{tabular}
\caption{Values of dimensionless exclusion volume $v_{\mbox{\scriptsize ex}}(K)/v(K)$ for selected shapes in dimensions 2 through 12. The aspect ratio of the cylinder and parallelpiped are respectively are $h/a=2$ and $h/b=2$ .}
\label{tab1}
\end{table}

\begin{table}[htp]
\centering
\begin{tabular}{l|@{\hspace{0.21cm}}c|@{\hspace{0.21cm}}c|@{\hspace{0.21cm}}c|@{\hspace{0.21cm}}c|@{\hspace{0.21cm}}c|@{\hspace{0.21cm}}c|@{\hspace{0.21cm}}c}
\hline\hline
 & $h/a=1$ & $h/a=2$  & $h/a=10$  & $h/a=50$  & $h/a=100$  & $h/a=500$  & $h/a=1000$  \\
\hline
$d$=2   & 4.12382 & 4.35657 & 6.75098 & 19.4307 & 35.3387 & 162.657 & 321.811 \\\hline
$d$=3   & 8.42857 & 9.2 & 16.8235 & 56.7013 & 106.684 & 506.67 & 1006.67  \\\hline
$d$=4   & 17.1691 & 19.205 & 38.781 & 140.393 & 267.683 & 1286.25 & 2559.48   \\\hline
$d$=5   & 34.9032 & 39.8261 & 86.2169 & 325.734 & 625.668 & 3025.61 & 6025.61  \\\hline
$d$=6   & 70.8532 & 82.2191 & 187.671 & 729.985 & 1408.92 & 6841.3 & 13631.9 \\\hline
$d$=7   & 143.672 & 169.176 & 402.869 & 1601.06 & 3100.82 & 15100.6 & 30100.6  \\\hline
$d$=8   & 291.067 & 347.206 & 856.329 & 3460.44 & 6719.46 & 32795 & 65389.9  \\\hline
$d$=9   & 589.233 & 711.097 & 1806.77 & 7400.04 & 14399.1 & 70398.4 & 140398  \\\hline
$d$=10  & 1192.07 & 1453.86 & 3790.24 & 15697.6 & 30596.4 & 149799 & 298805  \\\hline
$d$=11  & 2410.32 & 2968.11 & 7914.57 & 33089.3 & 64586 & 316583 & 631583  \\\hline
$d$=12  & 4871.15 & 6051.91 & 16464.4 & 69395.1 & 135613 & 665405 & 1327650 \\\hline
\hline
\end{tabular}
\caption{Values of dimensionless exclusion volume $v_{\mbox{\scriptsize ex}}(K)/v(K)$ for spherocylinders with selected aspect ratios $h/a$ in dimensions 2 through 12.}
\label{tab2}
\end{table}

\begin{table}[htp]
\centering
\begin{tabular}{l|@{\hspace{0.2cm}}c|@{\hspace{0.2cm}}c}
\hline\hline
 & Spherical Hyperplate & Cubical Hyperplate \\
\hline
$d$=3   & 2.35619 & 2.65868  \\\hline
$d$=4   & 5.54869 & 7.58925   \\\hline
$d$=5   & 12.1491 & 20.9595  \\\hline
$d$=6   & 25.7115 & 57.3242 \\\hline
$d$=7   & 53.4209 & 156.398  \\\hline
$d$=8   & 109.785 & 426.778  \\\hline
$d$=9  & 224.059 & 1165.98  \\\hline
$d$=10  & 455.162 & 3190.45  \\\hline
$d$=11  & 921.644 & 8744.25  \\\hline
$d$=12 & 1861.86 & 24004.1 \\\hline
\hline
\end{tabular}
\caption{Values of dimensionless exclusion volume $v_{\mbox{\scriptsize ex}}(K_H)/v_{\mbox{\scriptsize eff}}(K_H)$ for spherical and cubical hyperplates in dimensions 3 through 12. }
\label{tab_hyperplates}
\end{table}

\begin{table}[htp]
\centering
\begin{tabular}{l|@{\hspace{0.1cm}}c}
\hline\hline
 & Oriented Simplex \\
\hline
$d$=2  & 4  \\\hline
$d$=3  & 8 \\\hline
$d$=4  & 16  \\\hline
$d$=5  & 32   \\\hline
$d$=6  & 64 \\\hline
$d$=7  & 128  \\\hline
$d$=8  & 256 \\\hline
$d$=9  & 512  \\\hline
$d$=10 & 1024  \\\hline
$d$=11 & 2048  \\\hline
$d$=12 & 4096  \\\hline
\hline
\end{tabular}
\caption{Values of dimensionless exclusion volume $v^0_{ex}(K)/v(K)$ for oriented simplices in dimensions 2 through 12.}
\label{tab_oriented_simplex}
\end{table}



\newpage

\end{document}